\documentclass[prd,aps,floatfix,twocolumn,showpacs,nofootinbib]{revtex4}

\usepackage[dvipdfm]{graphicx}
\usepackage{bm}
\usepackage{amsmath,amssymb}
\usepackage{mathrsfs}
\usepackage{natbib}

\input{colordvi.tex}

\newcommand{\simgt}{\lower.5ex\hbox{$\; \buildrel > \over \sim \;$}}
\newcommand{\simlt}{\lower.5ex\hbox{$\; \buildrel < \over \sim \;$}}

\def\pd{{\rm d}}

\def\reff@jnl#1{{\rm#1\/}}
\def\aj{\reff@jnl{AJ}}                  
\def\araa{\reff@jnl{ARA\&A}}            
\def\apj{\reff@jnl{ApJ}}                
\def\apjl{\reff@jnl{ApJ}}               
\def\apjs{\reff@jnl{ApJS}}              
\def\ao{\reff@jnl{Appl.Optics}}         
\def\apss{\reff@jnl{Ap\&SS}}            
\def\aap{\reff@jnl{A\&A}}               
\def\aapr{\reff@jnl{A\&A~Rev.}}         
\def\aaps{\reff@jnl{A\&AS}}             
\def\azh{\reff@jnl{AZh}}                
\def\baas{\reff@jnl{BAAS}}              
\def\jcap{\reff@jnl{JCAP}}          
\def\jrasc{\reff@jnl{JRASC}}            
\def\memras{\reff@jnl{MmRAS}}           
\def\mnras{\reff@jnl{MNRAS}}            
\def\newastro{\reff@jnl{New Astron.}}   
\def\pra{\reff@jnl{Phys.Rev.A}}         
\def\prb{\reff@jnl{Phys.Rev.B}}         
\def\prc{\reff@jnl{Phys.Rev.C}}         
\def\prd{\reff@jnl{Phys.Rev.D}}         
\def\prl{\reff@jnl{Phys.Rev.Lett}}      
\def\pasp{\reff@jnl{PASP}}              
\def\pasj{\reff@jnl{PASJ}}              
\def\qjras{\reff@jnl{QJRAS}}            
\def\skytel{\reff@jnl{S\&T}}            
\def\solphys{\reff@jnl{Solar~Phys.}}    
\def\sovast{\reff@jnl{Soviet~Ast.}}     
\def\ssr{\reff@jnl{Space~Sci.Rev.}}     
\def\zap{\reff@jnl{ZAp}}                
\def\nat{\reff@jnl{Nature}}             
\def\physrep{\reff@jnl{Phys.~Rep.}}     
\def\prog{\reff@jnl{PThPS}}        

\begin{document}

\title{Nonlinear Biasing and Redshift-Space Distortions in Lagrangian Resummation Theory and N-body Simulations}

\author{Masanori Sato$^{1}$\footnote{masanori@a.phys.nagoya-u.ac.jp} and
Takahiko Matsubara$^{1,2}$}
\affiliation{%
$^{1}$ Department of Physics, Nagoya University, Nagoya 464--8602, Japan
}%
\affiliation{%
$^{2}$ Kobayashi-Maskawa Institute for the Origin of
 Particles and the Universe, Nagoya University, Nagoya 464--8602, Japan
}%

\date{\today}


\begin{abstract}
    Understanding a behavior of galaxy biasing is crucial for future
    galaxy redshift surveys. One aim is to measure
    the baryon acoustic oscillations (BAOs) within the precision of a
    few percent level. Using 30 large cosmological $N$-body
    simulations for a standard $\Lambda$CDM cosmology, we study the
    halo biasing over a wide redshift range. We compare the simulation
    results with theoretical predictions proposed by Matsubara
    which naturally incorporate the halo
    bias and redshift-space distortions into their formalism of
    perturbation theory with a resummation technique via the Lagrangian
    picture. The power spectrum and correlation function of halos
    obtained from
    Lagrangian resummation theory (LRT) well agree with $N$-body
    simulation results on scales of BAOs. Especially nonlinear effects
    on the baryon acoustic peak of the halo correlation function are
    accurately explained both in real and redshift space. We find that
    nonlinearity and scale dependence of bias are fairly well
    reproduced by 1-loop LRT up to $k=0.35h$Mpc$^{-1}$
    ($z=2$ and 3) within a few percent level in real space and up to
    $k=0.1h$Mpc$^{-1}$ ($z=2$) and 0.15$h$Mpc$^{-1}$ ($z=$3) in redshift space.
    Thus, the LRT is very powerful for accurately
      extracting cosmological information in upcoming high redshift
    BAO surveys.
\end{abstract}

\maketitle

\section{Introduction}
The redshift survey of galaxies is one of the most important methods
to probe the nature of a mysterious energy component called dark energy,
which is supposed to explain the late-time cosmic acceleration
discovered by the observation of distant
supernovae~\citep{1998AJ....116.1009R,1999ApJ...517..565P}. Baryon
acoustic oscillations (BAOs)
imprinted on the large scale structure can be used as a standard ruler
to measure the cosmic expansion history of the
Universe~\citep[e.g.,][]{2003ApJ...594..665B,2003PhRvD..68f3004H,2003ApJ...598..720S,2005ApJ...631....1G,2007ApJ...665...14S}.
The characteristic scale of BAOs, which is determined by the sound
horizon scale of baryon-photon plasma at the recombination epoch, is
though to be a robust
measure~\citep[e.g.,][]{1998ApJ...504L..57E,2001ApJ...557L...7C,2005ASPC..339..215H}.
Using the BAOs as a standard ruler, large galaxy surveys are expected
to provide a tight constraint on the nature of dark
energy~\citep[e.g.,][]{2003PhRvD..68h3504L,2004ApJ...615..573M,2005MNRAS.357..429A,2006astro.ph..9591A,2006MNRAS.366..884D,2009arXiv0901.0721A}.

That is why most of the planned galaxy redshift surveys aim
at measuring BAOs within the precision of a few percent level.
Some examples are BigBOSS~\citep{2009arXiv0904.0468S},
Euclid~\citep{2010arXiv1001.0061R}, the Hobby-Eberly Telescope Dark Energy
Experiment (HETDEX)\footnote{http://hetdex.org/}, the Large Sky Area
Multi-Object Fiber Spectroscopic Telescope
(LAMOST)\footnote{http://www.lamost.org/website/en}, the Square Kilometre
Array (SKA)\footnote{http://www.skatelescope.org/}, the Subaru Measurement
of Images and Redshifts (SuMIRe)\footnote{http://sumire.ipmu.jp/en/} and
the Wide-Field Infrared Survey Telescope
(WFIRST)\footnote{http://wfirst.gsfc.nasa.gov/}.
Recent observations of BAOs in modern galaxy surveys work well and
constrain the cosmological parameters with approximately 10\%
level~\citep[see, e.g.,][]{2005MNRAS.362..505C,2005ApJ...633..560E,2006PhRvD..74l3507T,2007MNRAS.378..852P,2007ApJ...657..645P,2008ApJ...676..889O,2010MNRAS.404...60R}.

However, in order to pursue an order-of-magnitude improvement, a
theoretically precise description of the BAOs is a crucial issue. It
needs to be investigated taking account of the various systematic
effects. The two main systematic effects are galaxy biasing and
redshift-space distortions. The spatial pattern of galaxy distribution
is not the same as that of dark matter and the galaxies are biased
tracers of mass~\citep{1984ApJ...284L...9K,1986ApJ...304...15B}. The
redshift of a galaxy does not purely reflect the Hubble flow, and
Doppler shift by a peculiar velocity is inevitably added. Therefore
the spatial pattern of clustering of galaxies is distorted in redshift
space~\citep{1987MNRAS.227....1K}. With time, these effects are
increasingly difficult to treat due to nonlinear evolution of
structure growth and degrade the contrast of the BAOs in the matter
power spectrum and/or correlation function, decreasing the
signal-to-noise ratio of the standard ruler method. The nonlinear
effects on BAOs play an important role in taking into account
percent-level cosmology although baryonic features appear on large
scales such as $100h^{-1}$Mpc. The amplitude of BAOs is increasingly
damped from small scales to large scales, with decreasing redshift
\citep[e.g.,][]{1999MNRAS.304..851M,2005ApJ...633..575S,2007ApJ...664..660E,2007APh....26..351H,2008ApJ...686...13S,2008PhRvD..77d3525S,2009PhRvD..80f3508P,2010ApJ...720.1650S,2011ApJ...734...94M}.
Redshift distortion effects enhance a nonlinear damping of BAOs along
the line-of-sight direction
\citep[e.g.,][]{1999MNRAS.304..851M,2005ApJ...633..575S,2007ApJ...664..660E,2007APh....26..351H,2008ApJ...686...13S,2008PhRvD..77d3525S,2009PhRvD..80f3508P,2010ApJ...720.1650S,2011ApJ...734...94M}.
Biasing is also affected by nonlinear effects and is scale
  dependent on BAOs scales. Such nonlinear effects impose a
serious problem in analyzing galaxy surveys
\citep[e.g.,][]{2006ApJ...645..977B,2007PhRvD..75f3512S,2007ApJ...657..645P,2008MNRAS.385..830S,2011arXiv1104.2948B}.

Since galaxies are expected to form within dark matter halos,
understanding and modeling the clustering properties of the halos
play an important role in model galaxy biasing. In the usual
halo model approach, the halo clustering is modeled by linear dynamics
and linear bias
factors~\citep{1996MNRAS.282..347M,2000MNRAS.318.1144P,2000MNRAS.318..203S,2002PhR...372....1C}.
Recently, \citet{2007PhRvD..75f3512S,2008PhRvD..78b3523S} and
\citet{2010arXiv1012.4833E} developed a nonlinear perturbation theory,
incorporating the effects of halo bias. In their approach,
there are some problems.
\citet{2007PhRvD..75f3512S,2008PhRvD..78b3523S} treat the halo bias as
a local bias in Eulerian space, although the halo bias is
intrinsically nonlocal in Eulerian space and does not fit well into
the local Eulerian biasing scheme~\citep{2011PhRvD..83h3518M}.
Meanwhile, the analytic approach developed by
\citet{2010arXiv1012.4833E} cannot deal with the redshift-space
distortions.

In this paper, we use nonlinear perturbation theory developed by
\citet{2008PhRvD..78h3519M} which naturally incorporate the halo bias
and redshift-space distortions in their formalism of
perturbation theory with a resummation technique through the
Lagrangian picture. A significant advantage of the Lagrangian
resummation theory (LRT) is that
  it is simpler and easier to calculate the power spectrum than other
resummation methods even in the presence of halo bias and
redshift-space distortions. The computational cost is similar to
that of standard perturbation theory
(SPT)~\citep[e.g.,][]{2002PhR...367....1B}. In this work, we
examine how well it reproduces the simulation results in both real and
redshift space. We focus not only on the halo power spectrum but also on
the halo correlation function, because cosmological information that can be
extracted from them is not exactly equivalent to each other because of
different error properties. In this paper, we use 1-loop LRT for
  simplicity. The 2-loop corrections should be useful to extend
  the valid range, once we confirm that the 1-loop results agree with
  simulations~\citep{2011arXiv1105.1491O}.

The structure of this paper is as follows. In
Section.~\ref{sec:analytical}, we briefly review analytical models
in order to obtain matter power spectrum in real and redshift space.
We also review the halo power spectrum based on 1-loop LRT. In
Section.~\ref{sec:Nbody} we describe the details of $N$-body simulation
and a method to calculate the power spectrum and two-point correlation
function from $N$-body simulations. In Section.~\ref{sec:result} we
show the main results. Section~\ref{sec:conc} is devoted to our
conclusions.

\section{Analytical Models}
\label{sec:analytical}
There are several analytical models to account for evolutions of
the matter power spectra in real and redshift space.
In this paper, we mainly use three analytical models to compare the
$N$-body simulation results; Linear perturbation theory
(LIN)~\citep[e.g.,][]{1993sfu..book.....P},
1-loop SPT~\citep[e.g.,][]{2002PhR...367....1B}, and
1-loop LRT~\citep[e.g.,][]{1989A&A...223....9B,1992MNRAS.254..729B,1993MNRAS.264..375B,1994MNRAS.267..811B,1995MNRAS.276..115C,1995A&A...298..643H,1997GReGr..29..733E,2004astro.ph.12025T,2008PhRvD..77f3530M}.

\subsection{Matter power spectra in real space}

Assuming that the amplitude of fluctuation is small, we can derive SPT
as expansion of the fluid equations. Schematically, the expansion is
written as 
\begin{equation}
 P_{\rm SPT}(k,z)=D_{+}(z)^2P_{\rm L,0}(k)+D_{+}(z)^4P_{\rm
   SPT,0}^{\mbox{\scriptsize 1-loop}}(k)+\cdots,\label{P_SPT_z}
\end{equation}
where $D_{+}(z)$ is the linear growth factor normalized to unity at
present and $P_{\rm L,0}(k)$ and $P_{\rm SPT,0}^{\mbox{\scriptsize
1-loop}}(k)$ are
the linear power spectrum and the 1-loop contribution to power spectrum,
respectively, at present. For the sake of convenience, by using
$P_{\rm L}(k,z)\equiv D_{+}(z)^2P_{\rm L,0}(k)$ and $P_{\rm SPT}^{\mbox{\scriptsize
  1-loop}}(k,z)\equiv D_{+}(z)^4P_{\rm SPT,0}^{\mbox{\scriptsize 1-loop}}(k)$, we
rewrite Eq.~(\ref{P_SPT_z}) as
\begin{equation}
 P_{\rm SPT}(k)=P_{\rm L}(k)+P_{\rm SPT}^{\mbox{\scriptsize 1-loop}}(k)+\cdots.\label{P_SPT}
\end{equation}
Here we drop the $z$ dependence. Throughout this paper, we do not
write the $z$ dependence unless otherwise stated.
The explicit expressions of the 1-loop contribution can be found in
the literature
\citep[e.g.,][]{1991PhRvL..66..264S,1992PhRvD..46..585M,1994ApJ...431..495J,2007PASJ...59.1049N,2009PhRvD..80d3531C}.

On the other hand, 1-loop LRT is written as \citep{2008PhRvD..77f3530M}
\begin{align}
 P_{\rm LRT}(k)=\exp\left(-k^2\sigma_{\rm v}^2\right)\left[P_{\rm
 SPT}(k)+k^2\sigma_{\rm v}^2P_{\rm L}(k)\right],
\label{P_LRT}
\end{align}
in terms of SPT. 
Here $\sigma_{\rm v}^2$ is the one-dimensional linear velocity
dispersion given by
\begin{equation}
 \sigma_{\rm v}^2=\frac{1}{3}\int\frac{\pd^3
  \vec{p}}{(2\pi)^3}\frac{P_{\rm L}(p)}{p^2}.
\label{vel_dis}
\end{equation}
If the exponential prefactor is expanded, the 1-loop SPT is
 exactly recovered 
when we consider the 1-loop level.

Finally we comment on other perturbation theories. Several other
approaches have been proposed beyond SPT, such as the renormalized
perturbation theory (RPT)
\citep{2006PhRvD..73f3519C,2006PhRvD..73f3520C,2008PhRvD..77b3533C},
the large-$N$ expansion \citep{2007A&A...465..725V}, the Time-RG method
\citep{2008JCAP...10..036P}, the renormalization group approach
\citep{2008MPLA...23...25M} and the closure theory
\citep{2008ApJ...674..617T}. In these newly developed approaches, the
standard perturbative expansion is reorganized and partially resummed
in various ways. Different levels of approximations and ansatz
  are used in those approaches.

\subsection{Matter power spectra in redshift space}
As in real space, we use SPT and LRT in redshift space.
Schematically, SPT in redshift space is written as
\begin{equation}
 P_{\rm SPT}^{\rm s}(k,\mu)=(1+f\mu^2)^2 P_{\rm L}(k)+P_{\rm SPT}^{\mbox{\scriptsize s,1-loop}}(k,\mu)+\cdots,\label{P_SPT_R}
\end{equation}
where $f=d\ln{D}/d\ln{a}$ is the logarithmic derivative of the linear
growth rate and $\mu$ is the cosine of the angle between the line of
sight and the wave vector $\vec{k}$. Here $P_{\rm
  SPT}^{\mbox{\scriptsize s,1-loop}}(k)$ is the 1-loop contribution
to the redshift-space power spectrum in SPT
\citep{2008PhRvD..77f3530M}. The first term on the right-hand side is
the linear-order result of the redshift-space power spectrum and the
factor $(1+f\mu^2)^2$ indicates the enhancement of the power spectrum
which is the so-called Kaiser effect \citep{1987MNRAS.227....1K}. The
Kaiser effect is represented as the coherent distortion by a peculiar
velocity along the line-of-sight direction.

On the other hand, 1-loop LRT in redshift space is written as \citep{2008PhRvD..77f3530M}
\begin{align}
 P_{\rm LRT}^{\rm
 s}(k,\mu)&=\exp\left[-k^2\sigma_{\rm v}^2[1+f(f+2)\mu^2]\right]\nonumber\\
 &\times\Bigl[(1+f\mu^2)^2 P_{\rm
 L}(k)+P_{\rm SPT}^{\mbox{\scriptsize s,1-loop}}(k,\mu)\nonumber\\
 &+(1+f\mu^2)^2[1+f(f+2)\mu^2]k^2\sigma_{\rm v}^2P_{\rm L}(k)\Bigr],\label{P_LRT_R}
\end{align}
As in the case in real space, if the exponential prefactor is
expanded, Eq.~(\ref{P_LRT_R}) reduces to the 1-loop SPT result.
The exponential prefactor shows the nonlinear damping effect by the
random motion of peculiar velocities, i.e., the Fingers-of-God effect
\citep{1972MNRAS.156P...1J,1977ApJ...212L...3S}.

We also use a model of redshift-space power spectrum, proposed
  by \citet{2004PhRvD..70h3007S}.
  This model gives
\begin{align}
 P_{\rm SCO}^{\rm
 s}(k,\mu)&=\exp\left(-f^2\mu^2k^2\sigma_{\rm
 v}^2\right)\nonumber\\
 &\times\left[P_{\delta\delta}(k)+2f\mu^2
 P_{\delta\theta}(k)+f^2\mu^4P_{\theta\theta}\right],\label{P_SCO}
\end{align}
where $P_{\delta\delta}(k)$, $P_{\theta\theta}(k)$ and
$P_{\delta\theta}(k)$ are auto power spectra of density and velocity,
and their cross power spectrum, respectively.
This model accounts for the nonlinear effects although it is still phenomenological.

In this paper, we consider only the monopole power spectrum which is
defined as 
\begin{equation}
 P^{\rm s}(k)=\frac{1}{2}\int_{-1}^{1}\pd{\mu}P^{\rm s}(k,\mu).
\end{equation}
Other multipole contributions to the power spectrum and correlation
function in redshift space, such as the quadrupole and hexadecapole are
examined in the literature
\citep[e.g.,][]{1998ASSL..231..185H,2006MNRAS.368...85T,2009PhRvD..80l3503T,2010PhRvD..82f3522T,2011arXiv1105.2037K,2011arXiv1105.1194K,2011arXiv1105.4165R,2011arXiv1103.3614T,2011PhRvD..83j3527T}.

\subsection{Halo power spectra in real and redshift space}
Based on 1-loop LRT, the power spectrum of the biased object in redshift space
is given by \citep{2008PhRvD..78h3519M}
\begin{align}
 P_{\rm hh, LRT}^{\rm
 s}(k,\mu)&=\exp\left[-k^2\sigma_{\rm
 v}^2[1+f(f+2)\mu^2]\right]\Bigl[(1+\langle{F'}\rangle\nonumber\\
&+f\mu^2)^2 P_{\rm
 L}(k)+\sum_{n,m}\mu^{2n}f^{m}E_{nm}(k)\Bigr],\label{P_HLRT_R}
\end{align}
where $\langle{F'}\rangle$ denotes a Lagrangian linear bias factor and
the explicit expression of $E_{nm}(k)$ is shown in \citet{2008PhRvD..78h3519M}.
$E_{nm}(k)$ includes the higher-order Lagrangian bias factor such as
$\langle{F''}\rangle$. 
Up to 1-loop order, we need only two bias factors
$\langle{F'}\rangle$ and $\langle{F''}\rangle$.
When we consider a mass range [$M_{1}$, $M_{2}$] of halos,
the bias function for the halo bias is given by \citep{2008PhRvD..78h3519M}
\begin{equation}
\langle{F^{(n)}}\rangle=\frac{(-1)^n}{\delta_c^n}\frac{\int_{M_1}^{M_2}\nu^n\frac{\pd^{n}g}{\pd\nu^n}\frac{\pd\ln\sigma}{\pd{M}}\frac{\pd{M}}{M}}{\int_{M_1}^{M_2}g(\nu)\frac{\pd\ln\sigma}{\pd{M}}\frac{\pd{M}}{M}},
\label{f_func}
\end{equation}
where $\delta_c$ is the critical overdensity at the present time. In an
Einstein-de Sitter cosmology, $\delta_c=1.686$. For general cosmology,
the value of $\delta_c$ shows weak dependence on cosmology
\citep{1993MNRAS.262..627L,1996MNRAS.282..263E,1997PThPh..97...49N,2000ApJ...534..565H}, so we include
cosmological dependence on $\delta_c$.
The quantity $\sigma$ is the root-mean-square linear density smoothed
with a top hat filter of radius $R$ and enclosing an average mass $M=\rho_0
4\pi{R}^3/3$,
\begin{equation}
\sigma^2(M)=\int\frac{k^2\pd{k}}{2\pi^2}W^2(kR)P_{\rm L}(k),
\end{equation}
with
\begin{equation}
W(x)=\frac{3}{x^3}(\sin x-x\cos x),
\end{equation}
where $\rho_0$ is the mean matter density of the Universe.
The quantity $\nu$ is defined by $\nu=\delta_c/\sigma$.
$g(\sigma)$ is the scaled differential mass function defined as \citep{2001MNRAS.321..372J}
\begin{equation}
 g(\sigma)=\frac{M}{\rho_0}\frac{\pd{n}}{\pd\ln\sigma^{-1}},
\end{equation}
where $n$ is the number density of halos with mass $M$.
The quantity $g(\sigma)$ is frequently used in the literature and there
have been several analytical predictions \citep{1974ApJ...187..425P,1991ApJ...379..440B,2001MNRAS.323....1S} and fitting formulae 
\citep[e.g.,][]{1999MNRAS.308..119S,2001MNRAS.321..372J,2006ApJ...646..881W,2007MNRAS.374....2R,2010MNRAS.403.1353C,2010MNRAS.402..589M,2011ApJ...732..122B}
for $g(\sigma)$.
In section \ref{sec:massfunc}, we will compare the mass function
obtained from our $N$-body simulations with several fitting formulae to
examine which fitting formulae are better.

From Eq.~(\ref{P_HLRT_R}) with substituting $f=0$, the power spectrum of
the biased object based on 1-loop LRT in real space is obtained as
\begin{align}
 P_{\rm hh, LRT}(k)=\exp\left[-k^2\sigma_{\rm
 v}^2\right]\left[(1+\langle{F'}\rangle)^2 P_{\rm L}(k)+E_{00}(k)\right].\label{P_HLRT}
\end{align}
By expanding the exponential prefactor and considering the linear term in
$P_{\rm L}(k)$, the linear result is derived as
\begin{align}
 P_{\rm hh, LIN}(k)=(1+\langle{F'}\rangle)^2 P_{\rm L}(k).\label{P_HLIN}
\end{align}
In linear theory, using the Eulerian linear bias factor $b$, the power
spectrum of the biased object is defined by
\begin{equation}
 P_{\rm hh, LIN}(k)=b^2 P_{\rm L}(k).\label{P_HLIN_O}
\end{equation}
Comparing Eq.~(\ref{P_HLIN}) and Eq.~(\ref{P_HLIN_O}), we can easily
derive the relation between Eulerian and Lagrangian biases as
\begin{equation}
 b=1+\langle{F'}\rangle.
\label{Euler_bias}
\end{equation}

This result is the same as that derived from the halo model approach by using
the spherical collapse model
\citep{1996MNRAS.282..347M,1997MNRAS.284..189M}, i.e., the Eulerian linear
bias is given by unity plus the Lagrangian linear bias~(see,
\citep{2011PhRvD..83h3518M} for more accurate expressions on nonlinear scales).
It should be noted that the result of Eq.~(\ref{Euler_bias}) is derived
without assuming a spherical collapse model.

The corresponding linear result in redshift space is obtained by a
linear term of Eq.~(\ref{P_HLRT_R}) as
\begin{equation}
 P_{\rm hh, LIN}^{\rm s}(k,\mu)=(1+\langle{F'}\rangle+f\mu^2)^2 P_{\rm L}(k).
\end{equation}
By using Eq.~(\ref{Euler_bias}), this equation is rewritten by
\begin{equation}
 P_{\rm hh, LIN}^{\rm s}(k,\mu)=b^2(1+\beta\mu^2)^2 P_{\rm L}(k),
\label{P_HLIN_R}
\end{equation}
where $\beta=f/b$ is the redshift-space distortion parameter. This
equation is equivalent to the Kaiser formula
\citep{1987MNRAS.227....1K}.
Methods used to determine the redshift-space distortion parameter
$\beta$ are summarized by \citet{1998ASSL..231..185H}.

\section{$N$-body Simulations}
\label{sec:Nbody}
\subsection{Simulation Parameters}

\begin{table*}
\caption{
Parameters used in low- and high-resolution $N$-body simulations. 
}
\label{table1}
\begin{center}
\begin{tabular}{p{35mm}ccccccccccc}
\hline  
Name & $\Omega_{\rm m}$ & $\Omega_{\Lambda}$ &
 $\Omega_{\rm b}$ & $h$ & $n_s$ & $\sigma_8$ & $L_{\rm box}$ & $N_p$ &
 $z_{\rm ini}$ & $r_{\rm s}$ & $N_{\rm run}$\\ \hline
L1000(low resolution) & 0.265 & 0.735 & 0.0448 & 0.71 & 0.963 & 0.80 & 1000$h^{-1}$Mpc &
				 1024$^3$ & 36 & 50$h^{-1}$kpc & 30\\ 
L500(high resolution) & 0.265 & 0.735 & 0.0448 & 0.71 & 0.963 & 0.80 & 500$h^{-1}$Mpc &
				 1024$^3$ & 42 & 25$h^{-1}$kpc & 5\\ \hline
\end{tabular}
\end{center}
\end{table*}

We use the numerical simulation code {\em
  Gadget2}~\citep{2005MNRAS.364.1105S} in its full tree-particle mesh
mode. For all of the simulations discussed in the present paper, we
adopted the standard $\Lambda$CDM model
with the matter density $\Omega_{m}=0.265$, baryon density
$\Omega_{b}=0.0448$, dark energy density $\Omega_{\Lambda}=0.735$ with
equation of state parameter $w=-1$, spectral index $n_s=0.963$, the
variance of the density fluctuation in a sphere of radius 8
$h^{-1}$Mpc $\sigma_8=0.80$, and Hubble parameter $h=0.71$. These
cosmological parameters are consistent with the Wilkinson Microwave
Anisotropy Probe 7yr (WMAP7) results~\citep{2011ApJS..192...18K}. We
employ $N_p=1024^3$ particles in boxes of side $L_{\rm
  box}=1000h^{-1}$Mpc with softening length $r_{\rm s}$ being
$50h^{-1}$kpc and of side $L_{\rm box}=500h^{-1}$Mpc with softening
length $r_{\rm s}$ being $25h^{-1}$kpc, respectively abbreviated to
L1000 and L500. Throughout this paper, we show results obtained from
L1000 unless otherwise stated. The initial conditions are
generated based on the 2nd-order Lagrangian perturbation theory
(2LPT)~\citep{2006MNRAS.373..369C,2009PASJ...61..321N} with the
initial linear power spectrum calculated by {\em
  CAMB}~\citep{2000ApJ...538..473L}. 
We use parallelized 2LPT code which is kindly provided by Takahiro
Nishimichi who developed it in \citet{2011A&A...527A..87V} to run
  large cosmological $N$-body simulations with initial conditions based
  on 2LPT.
The initial redshift is set to
$z_{\rm ini}=36$ for L1000 and $z_{\rm ini}=42$ for L500. We perform
$N_{\rm run}=30$ realizations for L1000 and $N_{\rm run}=5$
realizations for L500. Table~\ref{table1} summarizes the parameters
used in the simulations.

\begin{table*}
\caption{
Properties of halo catalogues for low- and high-resolution simulations.
 $\bar{N}_h$, $\bar{n}_h$ and $\bar{M}_h$
 are the ensemble average number, number densities and mass of halos
 for redshifts we employed. 
}
\label{table2}
\begin{center}
\begin{tabular}[c]{p{15mm}ccc|p{15mm}ccc}
\multicolumn{4}{c|}{L1000(low resolution)}&
\multicolumn{4}{c}{L500(high resolution)}\\
\hline  
  & $\bar{N}_h$ & $\bar{n}_h[h^3{\rm Mpc}^{-3}]$ &
 $\bar{M}_h[h^{-1}M_{\odot}]$ & & $\bar{N}_h$ & $\bar{n}_h[h^3{\rm
 Mpc}^{-3}]$ & $\bar{M}_h[h^{-1}M_{\odot}]$ \\ \hline
$z=3$ & 4.00$\times{10}^5$ & 4.00$\times{10}^{-4}$ &
 2.59$\times{10}^{12}$ & $z=3$ & 1.08$\times{10}^6$ & 8.68$\times{10}^{-3}$ &
 4.58$\times{10}^{11}$\\ 
$z=2$ & 1.21$\times{10}^6$ & 1.21$\times{10}^{-3}$ &
 3.30$\times{10}^{12}$ & $z=2$ & 1.86$\times{10}^6$ & 1.48$\times{10}^{-2}$ &
 6.12$\times{10}^{11}$\\
$z=1$ & 2.38$\times{10}^6$ & 2.38$\times{10}^{-3}$ &
 4.75$\times{10}^{12}$ & $z=1$ & 2.42$\times{10}^6$ & 1.94$\times{10}^{-2}$ &
 9.07$\times{10}^{11}$\\
$z=0.5$ & 2.82$\times{10}^6$ & 2.82$\times{10}^{-3}$ &
 5.99$\times{10}^{12}$ & $z=0.5$ & 2.52$\times{10}^6$ & 2.01$\times{10}^{-2}$ &
 1.15$\times{10}^{12}$\\
$z=0.3$ & 2.93$\times{10}^6$ & 2.93$\times{10}^{-3}$ &
 6.63$\times{10}^{12}$ & $z=0.3$ & 2.52$\times{10}^6$ & 2.01$\times{10}^{-2}$ &
 1.27$\times{10}^{12}$\\
$z=0$ & 3.05$\times{10}^6$ & 3.05$\times{10}^{-3}$ &
 7.73$\times{10}^{12}$ & $z=0$ & 2.49$\times{10}^6$ & 1.99$\times{10}^{-2}$ &
 1.47$\times{10}^{12}$\\\hline
\end{tabular}
\end{center}
\end{table*}

We store outputs at $z=3$, 2, 1, 0.5, 0.3, and 0 and identify halos for
each output using a friends-of-friends (FOF) group finder with linking length
of 0.2 times the mean separation~\citep{1985ApJ...292..371D}. 
We select halos in which the number of particles, $N_p$, is equal to or
larger than 20 which corresponds to the halos with masses $1.37\times
10^{12}h^{-1}M_{\odot}$ for L1000 and $1.71\times
10^{11}h^{-1}M_{\odot}$ for L500.
The average number, number densities, and mass of halos among realizations for
redshifts at which we store outputs can be found in Table~\ref{table2}.

\subsection{Analysis: Power Spectra and Two-Point Correlation Functions}
To calculate the power spectrum from $N$-body simulations, we calculate 
the Fourier transform of the density field, denoted as
$\Tilde{\delta}^n(\vec{k})$, where the superscript $n$ denotes the $n$-th
realization and $\vec{k}$ shows the wave number. 
First, we assign the $N$-body particles onto
$N_{\rm grid}^3=1024^3$ grids based on the cloud-in-cell (CIC) mass assignment
scheme~\citep{1988csup.book.....H}. 
We then use Fast Fourier Transformation (FFT) to calculate the
density contrast in Fourier space and correct the
effect of the CIC mass assignment scheme as~\citep{2008MNRAS.383..755A,2009ApJ...700..479T}
\begin{equation}
 \Tilde{\delta}^n(\vec{k})\rightarrow\frac{\Tilde{\delta}^n(\vec{k})}{
  \left[{\rm sinc}\left(\frac{k_x L_{\rm box}}{2N_{\rm grid}}\right)
  {\rm sinc}\left(\frac{k_y L_{\rm box}}{2N_{\rm grid}}\right)
  {\rm sinc}\left(\frac{k_z L_{\rm box}}{2N_{\rm grid}}\right)\right]^2},
\label{corr_del}
\end{equation}
where ${\rm sinc}(x)=\sin(x)/x$. A more careful analysis was done by
\citet{2005ApJ...620..559J}. Finally, squaring the density contrast in
Fourier space and taking an average over Fourier modes and
realizations, the ensemble average of binned power spectrum is
given by
\begin{align}
 \hat{P}(k_i)&=\frac{1}{N_{\rm run}N_i^k}\sum_{n=1}^{N_{\rm
  run}}\sum_{k_i^{\rm min}<\lvert\vec{k}\rvert <k_i^{\rm
 max}}\left\lvert\Tilde{\delta}^n(\vec{k})\right\rvert^2,\label{def_pk}\\
 k_i&\equiv\frac{1}{N_i^k}\sum_{k_i^{\rm min}<\lvert\vec{k}\rvert <k_i^{\rm
 max}}\lvert\vec{k}\rvert,
\label{def_ki}
\end{align}
where $N_i^k$ is the number of Fourier modes in the $i$-th wave number
bin, and $k_i^{\rm min}$ and $k_i^{\rm max}$ are the minimum and
maximum wave number of the $i$-th bin, respectively. The obtained
power spectrum is contaminated by the shot noise effect due to the
discreteness of density field in $N$-body simulations. We simply
assume the Poisson model where the shot noise is given by the inverse
of number density of dark matter particles, $1/\bar{n}=L_{\rm
  box}^3/N_p$ and then subtract the shot noise, though this effect is
very small and only has an impact on the result at $z=3$ on
scales we have considered. Note that the power spectra measured from
the treatment above suffer from the effect of finite-mode sampling due
to a finite number of realizations or a finite box size, shown by
\citet{2008MNRAS.389.1675T}. The finite-mode effect is important at
$k\simlt 0.1h{\rm Mpc^{-1}}$. We do not eliminate finite-mode effect
because extension of their method to the halo case we are
interested in is not trivial. Hence, to overcome this effect we run
many realizations, i.e., $N_{\rm run}=30$ and use a large simulation box
size, i.e., $L_{\rm box}=1000h^{-1}{\rm Mpc}$.

To calculate the two-point correlation function from $N$-body
simulations, we adopt the grid-based calculation using FFT.
\citet{2009PhRvD..80l3503T} show a grid-based calculation with FFT almost
coincides with the direct pair-count method (see, Appendix C in their paper).
Since the grid-based calculation with FFT is computationally less expensive
than the direct pair counting, we therefore adopt the grid-based
calculation.
It should be noted that the grid-based calculation is limited to scales
larger than the grid size $r>L_{\rm box}/N_{\rm grid}$.
In this method, we first compute the square of the density field in
Fourier space on each grid.
Then taking the inverse Fourier transformation and an average over
distances and realizations, we can obtain the two-point correlation
functions. This is expressed as \citep{2009PhRvD..80l3503T}
\begin{equation}
 \hat{\xi}(r_i)=\frac{1}{N_{\rm run}N_i^r}\sum_{n=1}^{N_{\rm
  run}}\sum_{r_i^{\rm min}<\lvert\vec{r}\rvert <r_i^{\rm
 max}}{\rm FFT}^{-1}\left[\left\lvert\Tilde{\delta}^n(\vec{k})\right\rvert^2;\vec{r}\right],
\end{equation}
where ${\rm FFT}^{-1}$ denotes the inverse FFT of the square of the
density field in Fourier space. Here, $N_i^r$ is the number of modes in
the $i$-th
distance bin, and $r_i^{\rm min}$ and $r_i^{\rm max}$ are the minimum
and maximum distances of the $i$-th bin, respectively.
We chose $r_i$ to be the center of the $i$-th bin, i.e., $r_i=(r_i^{\rm
min}+r_i^{\rm max})/2$.

For the estimation of the halo power spectrum, we adopt the same method
used for the power spectrum estimation to the halo cases, but 
apply the FOF halo mass correction as will be discussed in
Section.~\ref{sec:massfunc}. To be more precise, we impose
Eq.~(\ref{fofcor}) to halo particles when we assign the halo particles
based on the CIC mass assignment scheme in order to introduce the FOF mass
correction for $N$-body simulations. Then replacing $\Tilde{\delta}^n$
with $\Tilde{\delta}_h^n$, we do the same calculation described above
[from Eqs.~(\ref{corr_del})-(\ref{def_ki})] in order to obtain the
halo power spectrum. $\Tilde{\delta}_h^n$ is the density field of
halos in Fourier space in the $n$-th realization.

There is one concern about shot noise. If the dark matter halos are
regarded as a Poisson sampling, the subtraction term is the
same as the dark matter particle case but using the appropriate number
density $\bar{n}_h$. However \citet{2007PhRvD..75f3512S} found that
this standard correction method is not exactly correct for halos,
particularly for those of large mass. This is probably because
in order to identify halos by using the FOF algorithm, we automatically
impose that distances between halos are larger than the sum of their
radii, or they would have been linked as bigger halos. To correctly
eliminate the shot noise, \citet{2007PhRvD..75f3512S} proposed a new
procedure that includes this exclusion effect (see, Appendix A in
their paper for more details and also see,
\citet{2009PhRvL.103i1303S,2010PhRvD..82d3515H} for a new method to
suppress the shot noise effect by weighting halos). However, we do
not use this procedure to subtract the shot noise, but use the standard
correction method for simplicity.
The result could not being changed on scales we are interested
in, because the number density of halos are large enough, and there
is little shot noise effect.


For the estimation of the halo two-point correlation function, we use
a grid-based calculation with FFT as is used in matter correlation function.

\section{Results}
\label{sec:result}
\subsection{Matter Power Spectra}

\begin{figure*}[!t]
\begin{minipage}{.48\textwidth}
\begin{center}
\includegraphics[width=0.95\textwidth]{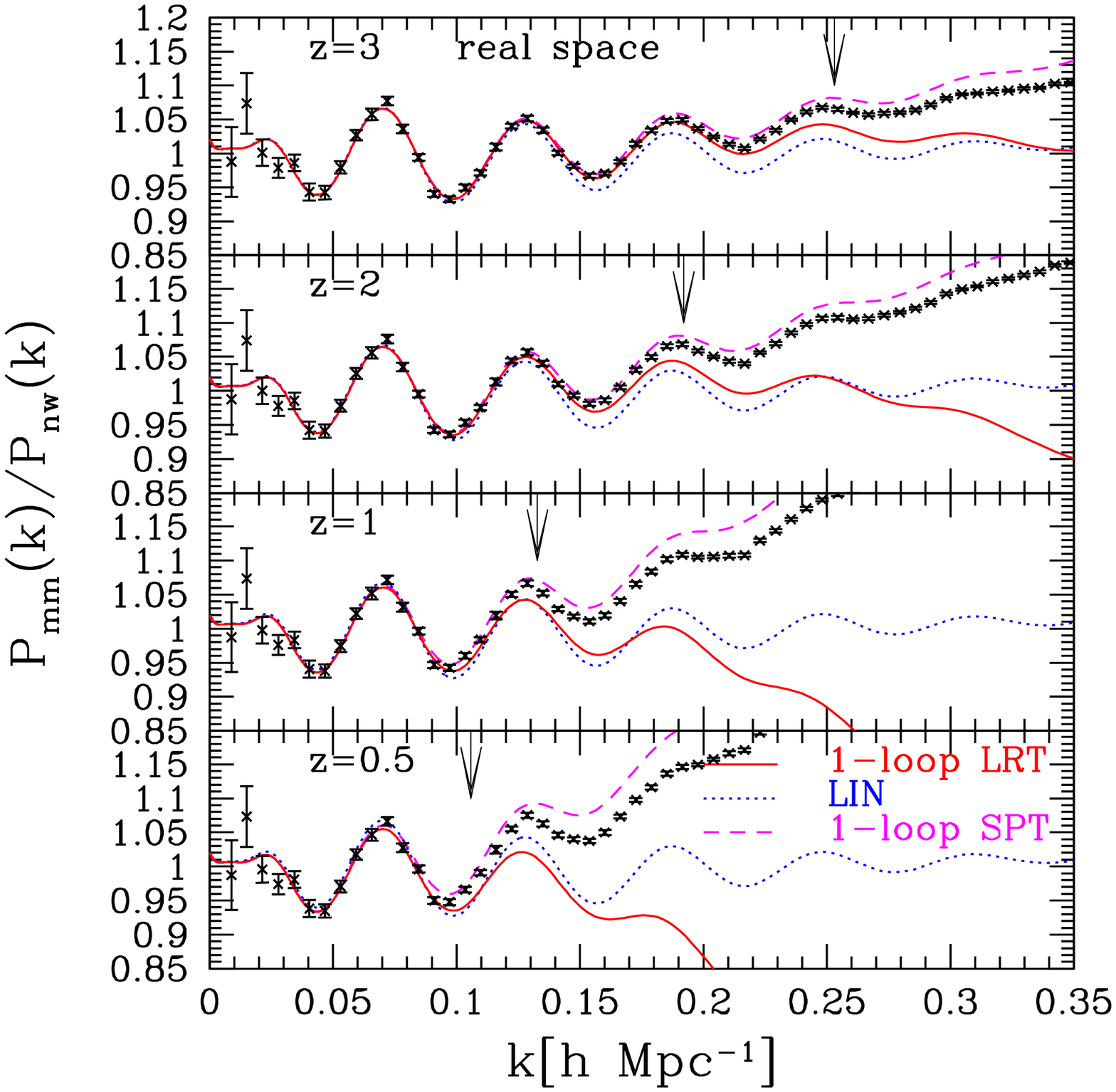}
\end{center}
\end{minipage}
\begin{minipage}{.48\textwidth}
\begin{center}
\includegraphics[width=0.95\textwidth]{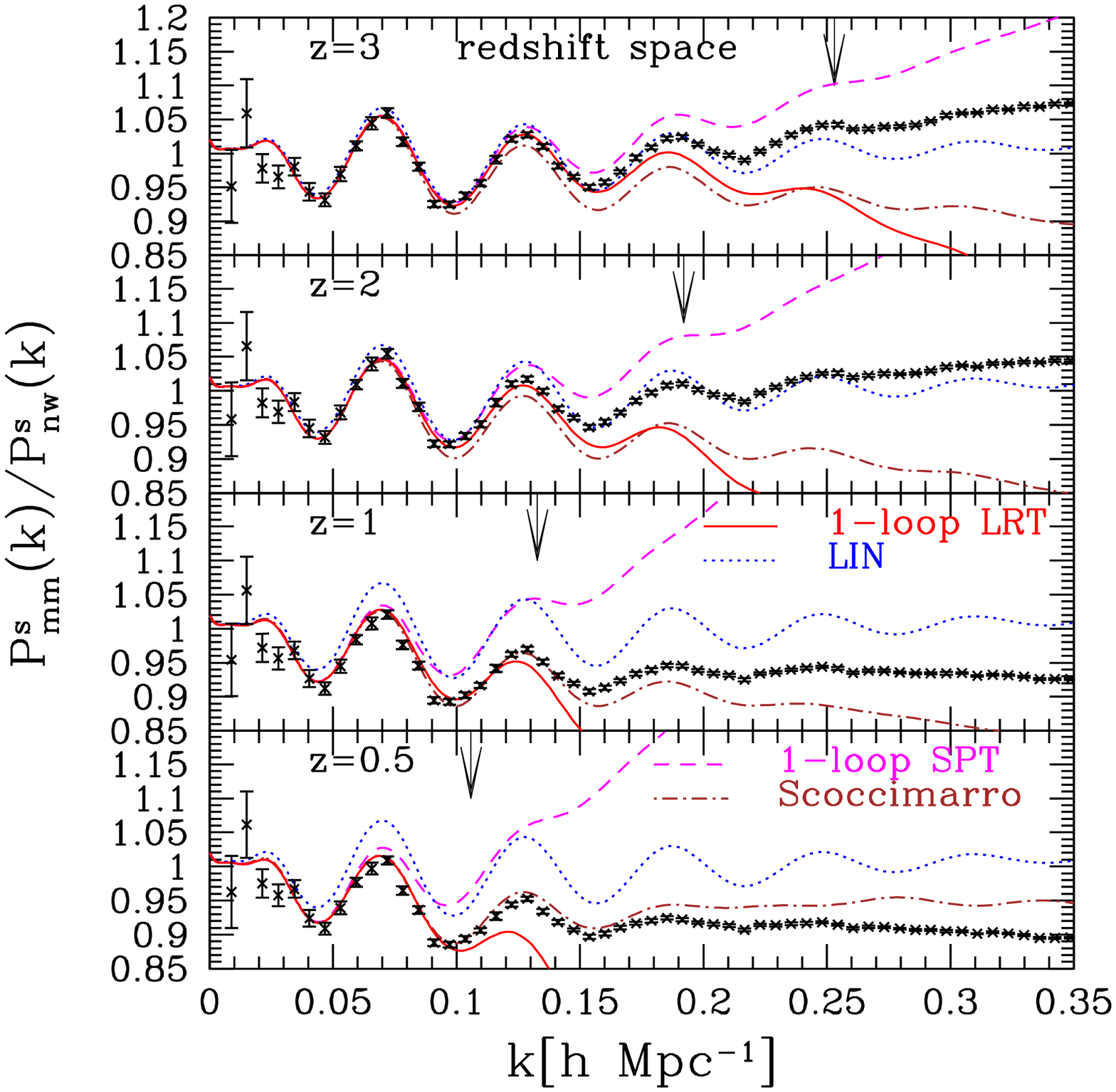}
\end{center}
\end{minipage}
\vskip-\lastskip
\caption{Comparison of mass power spectra obtained from $N$-body
 simulations to analytical predictions in real (left panel) and redshift
 (right panel)
 space for redshifts, $z=3$, 2, 1, and 0.5 from top to bottom.
 Power spectra are normalized by the no-wiggle fitting formula $P_{\rm
 nw}(k)$ \citep{1998ApJ...496..605E} in real space and by the no-wiggle
 fitting formula taking the Kaiser effect into account $P^s_{\rm nw}(k)$ in
 redshift space, respectively.
 The solid, dotted and dashed curves show the predictions of the 1-loop
 Lagrangian resummation theory (1-loop LRT), the linear theory (LIN),
 and the 1-loop standard perturbation theory (1-loop SPT), respectively.
 The dot-dashed curves in the right panel show
 the predictions of the empirical model proposed by \citet{2004PhRvD..70h3007S}.
 The arrows indicate the valid ranges for several redshifts, where
 the results of 1-loop LRT are expected to be accurate within a few percent.
}
\label{fig:matter_pk}
\end{figure*}

Figure~\ref{fig:matter_pk} shows the matter power spectra obtained from
$N$-body simulations among 30 realizations for $z=3$, 2, 1 and 0.5.
We compare them with several analytical predictions. The left and right
panels show the results of power spectrum divided by the smooth linear
power spectrum $P_{\rm nw}(k)$ which is calculated by using the no-wiggle
fitting formula of \citet{1998ApJ...496..605E} in real
space and by the no-wiggle power spectrum taking the Kaiser effect
($1+2f/3+f^2/5$) into account in redshift space, respectively.

All of the theoretical predictions well agree with the $N$-body simulations
within a range of the error bars on large scales.
The range of agreement in all of the theoretical predictions is generally wider
as redshift is higher.
This is because the amplitude of density fluctuation is smaller at
higher redshift.
As has been stated in the literature
\citep{2009PhRvD..80d3531C,2009PASJ...61..321N}, the result of 1-loop SPT
(dashed curves) is not sufficiently accurate to describe the BAOs not
only in real space but also in redshift space.
It seems that the discrepancy in redshift space
between the 1-loop SPT prediction and simulation results is larger than that in
real space. 
The amplitude of the power spectrum from 1-loop SPT generally
overestimates that from $N$-body simulations.
However, the range of the agreement in 1-loop SPT is wider in both real and
redshift space than that in LIN (dotted curves), because 1-loop SPT includes
the next-order contribution of nonlinear growth.
Meanwhile, the amplitude of the power spectrum in 1-loop LRT (solid
curves) rapidly falls off at a certain wave number and it deviates from
the $N$-body results. This is attributed to the exponential prefactor in
Eqs.~(\ref{P_LRT}) and (\ref{P_LRT_R}). In real space, the agreement of
1-loop LRT is roughly equivalent to
that of 1-loop SPT. However, we can see that the 1-loop LRT results give a
better agreement
with simulations in redshift space. We also plot another theoretical
prediction proposed by \citet{2004PhRvD..70h3007S} in redshift space
which is denoted as the dot-dashed curves in the right panel of
Figure~\ref{fig:matter_pk}.
The model of \citet{2004PhRvD..70h3007S} seems to be a good agreement
at certain redshift ($z=0.5$). However, this is just coincidence
because it does not describe the simulation results at different
redshifts.

Comparing the power spectra in real and redshift space, the
amplitude of the power spectra on small scales in redshift space is
suppressed by nonlinear redshift-space distortions. This suppression
is due to the random motion of peculiar velocities in virialized
objects which is known as the Fingers-of-God effect
\citep{1972MNRAS.156P...1J,1977ApJ...212L...3S}.
 
The arrows in Figure~\ref{fig:matter_pk} indicate the reliable ranges
$k_{\rm NL}/2$ for several redshifts, where the results of 1-loop LRT are
expected to be accurate within a few percent. This criterion is
proposed by \citet{2008PhRvD..77f3530M}. It is determined by the
damping scale of the exponent which is estimated from
Eq.~(\ref{vel_dis}) as
\begin{equation}
 k_{\rm NL}\equiv\frac{1}{\sigma_{\rm
  v}}=\left[\frac{1}{6\pi^2}\int\pd{p}{P_{\rm L}}(p)\right]^{-1/2}.
\end{equation}

\subsection{Correlation Functions}

\begin{figure*}[!t]
\begin{minipage}{.48\textwidth}
\begin{center}
\includegraphics[width=0.95\textwidth]{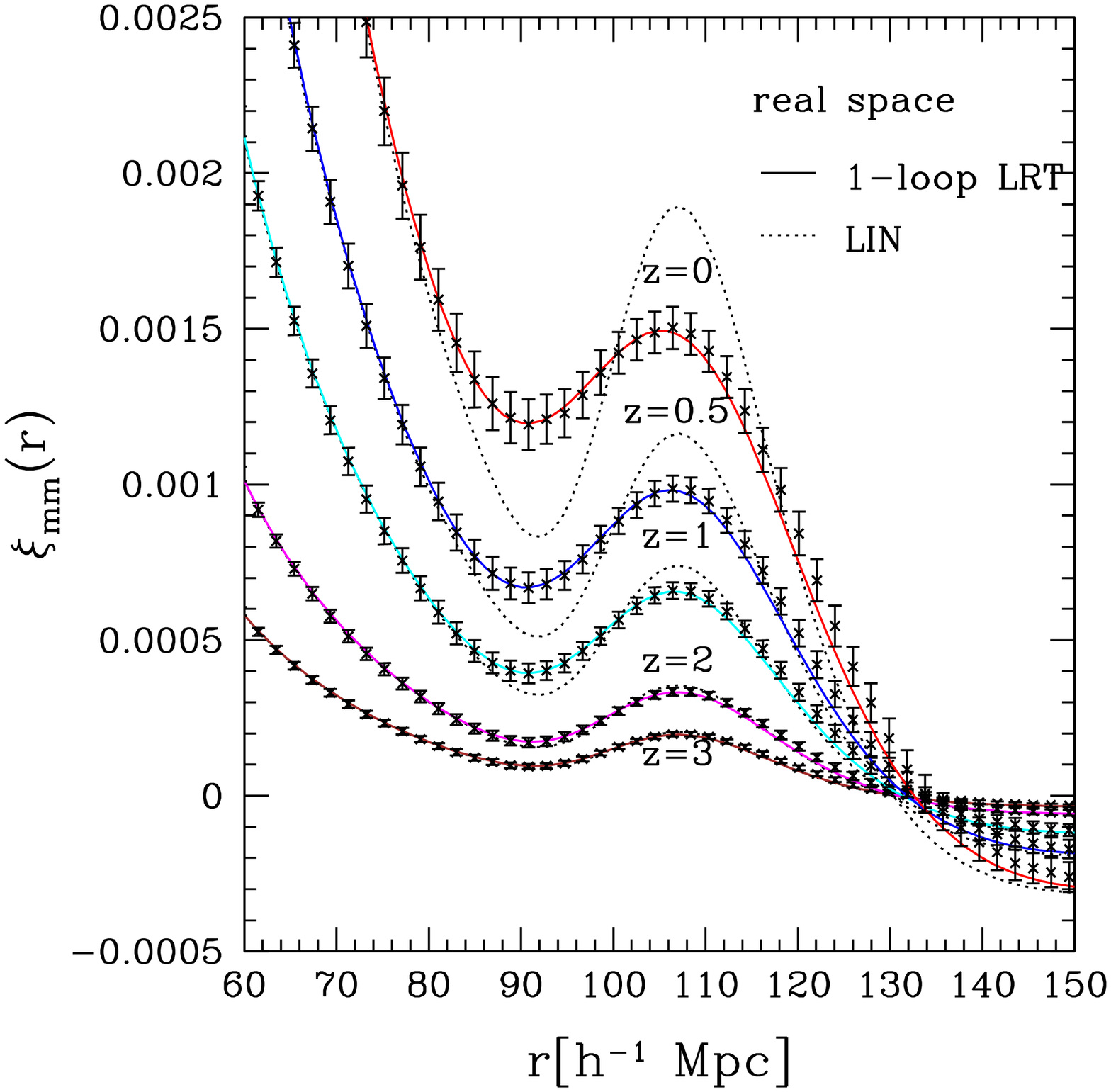}
\end{center}
\end{minipage}
\begin{minipage}{.48\textwidth}
\begin{center}
\includegraphics[width=0.95\textwidth]{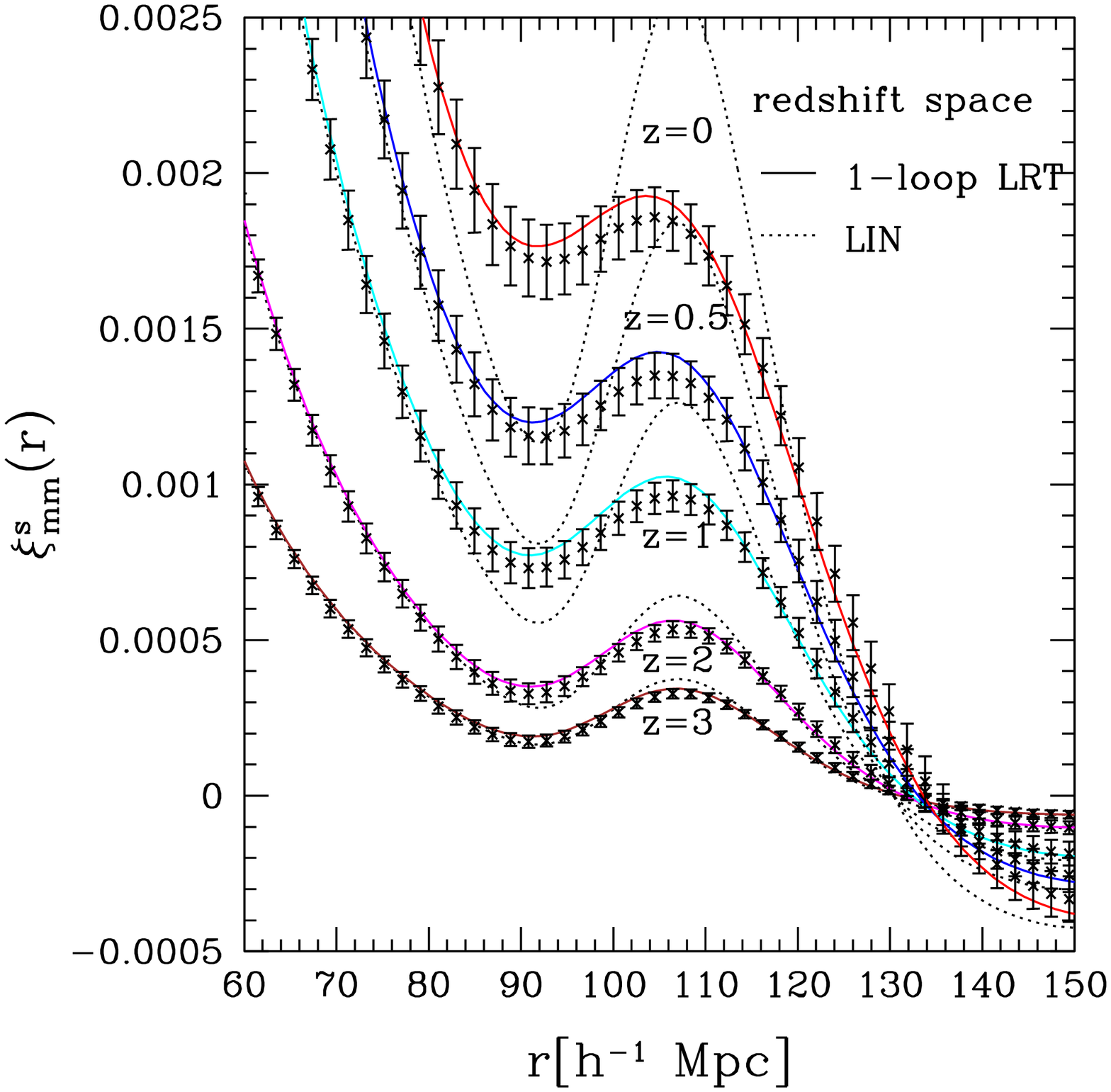}
\end{center}
\end{minipage}
\vskip-\lastskip
\caption{Comparison of two-point correlation functions obtained from
 $N$-body simulations to analytical predictions in real (left panel) and
 redshift (right panel) space for redshifts, $z=3$, 2, 1, 0.5, and 0 from
 bottom to top. The solid curves represent the results of the 1-loop Lagrangian
 resummation theory while the dotted curves show the results of the linear
 theory.
}
\label{fig:matter_xi}
\end{figure*}

Next we focus on the two-point correlation function. Although
the correlation function and power spectrum are directly related by
Fourier transforms and have mathematically equivalent information,
cosmological information that can be extracted from them with real
data is not exactly equivalent because error properties are different.
Therefore it is important to examine not only the power spectrum but
also the two-point correlation function. Two-point correlation
functions in real and redshift space around baryon acoustic peaks at
$z=3$, 2, 1, 0.5, and 0 are shown in left and right panels of
Figure~\ref{fig:matter_xi}, respectively. The theoretical two-point
correlation function can be expressed in terms of the power spectrum as
\begin{equation}
 \xi(r)=\int\frac{{k^2}\pd{k}}{2\pi^2}\frac{\sin(kr)}{kr}P(k).
\label{twopt}
\end{equation}
The $N$-body results clearly deviate from the LIN prediction (dotted curves) as
decreasing the redshift, because nonlinear growth of the structure becomes
significant as decreasing the redshift. In contrast, 1-loop LRT results
(solid curves) fairly well reproduce the $N$-body results at all the redshift
we have considered in both real and redshift space, although in
redshift space a closer look around acoustic peaks reveals that the
$N$-body results deviate from 1-loop LRT results as decreasing the redshift.
Although the 1-loop LRT predictions considerably deviate from $N$-body
  simulation results on large wave number in the power spectrum
  (Figure~\ref{fig:matter_pk}), the acoustic peak structure in
the correlation function comes from the low-$k$ behavior of the power
spectrum, and the power spectrum at low-$k$ is accurately described by
1-loop LRT \citep{2008PhRvD..77f3530M,2008MNRAS.390.1470S}.

The baryon acoustic peaks in both real and redshift space tend to be smeared as
decreasing the redshift. However, the effect seems stronger than those in
real space due to the Fingers-of-God effect.
In the power spectrum results (Figure~\ref{fig:matter_pk}), the difference
between 1-loop SPT and 1-loop LRT seems to be small, especially in real
space. In correlation function results, however, the difference is quite
big, because SPT cannot predict the correlation function which is
ascribed to fail to convergence the integral in Eq.~(\ref{twopt})
\citep{2008PhRvD..77f3530M}.
Therefore, we conclude that the LRT is very powerful and sufficient to
predict the BAOs in correlation functions.

\subsection{Halo Mass Functions}
\label{sec:massfunc}

\begin{figure*}[!t]
\begin{minipage}{.48\textwidth}
\begin{center}
\includegraphics[width=0.95\textwidth]{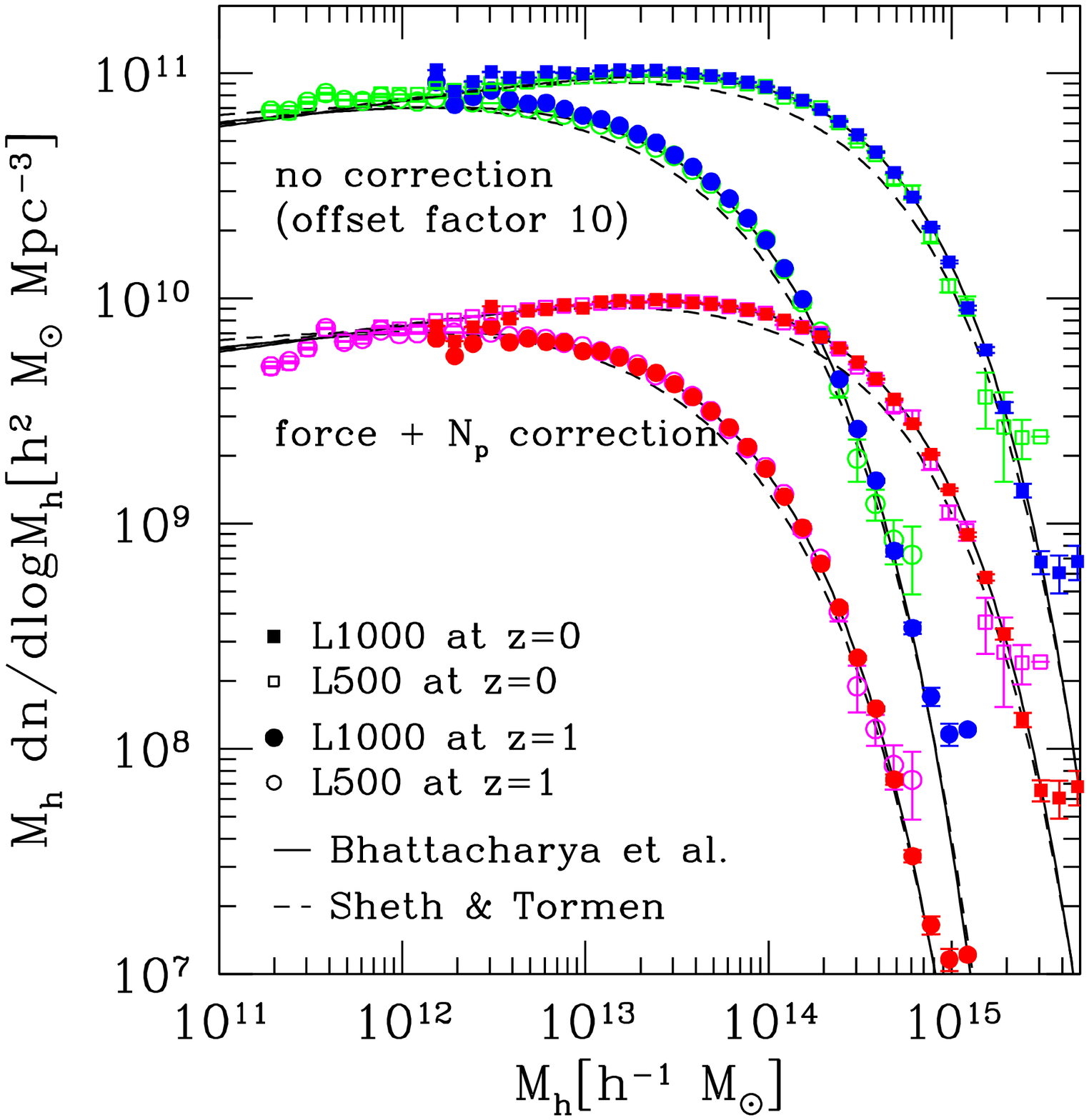}
\end{center}
\end{minipage}
\begin{minipage}{.48\textwidth}
\begin{center}
\includegraphics[width=0.95\textwidth]{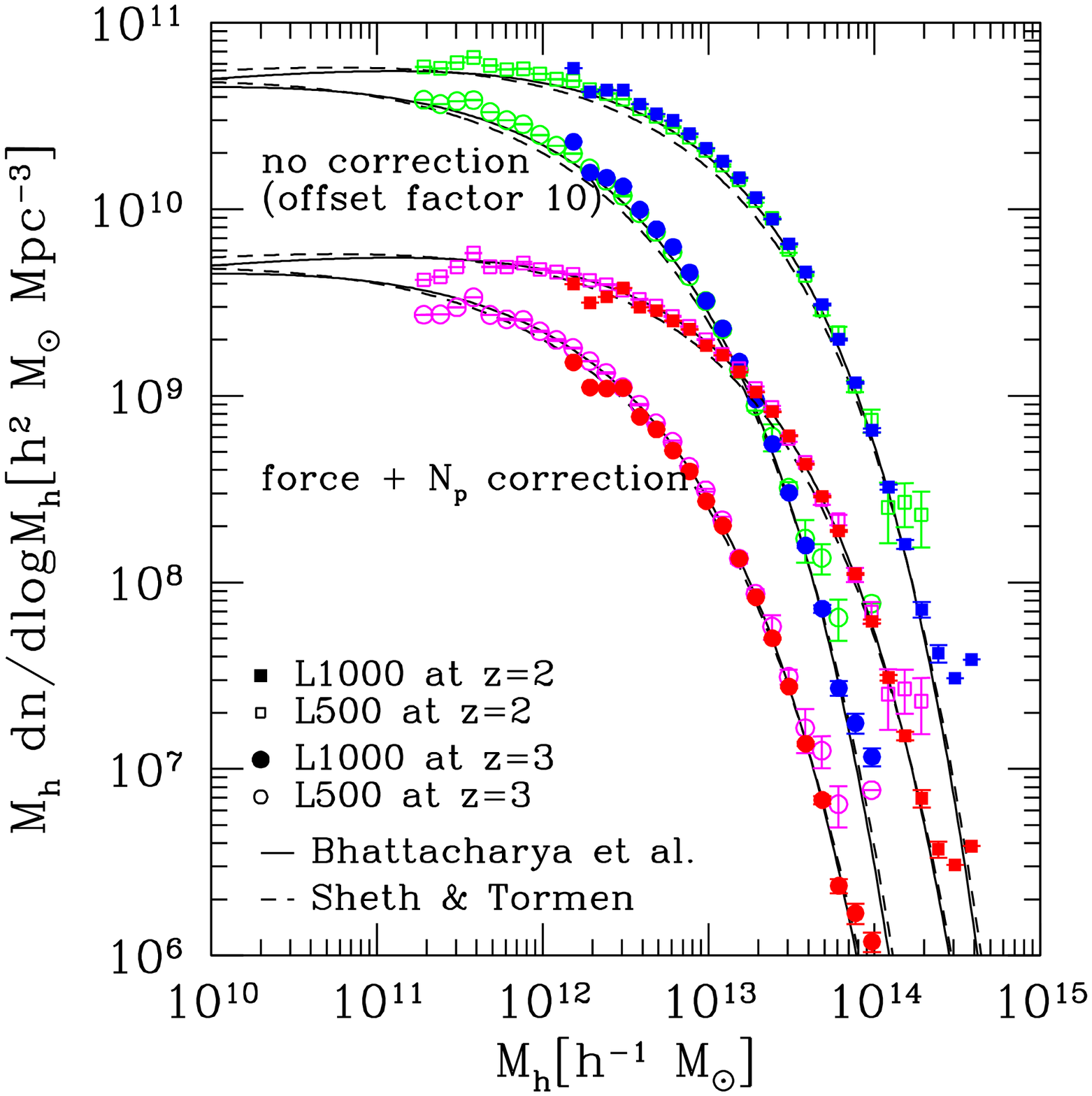}
\end{center}
\end{minipage}
\vskip-\lastskip
\caption{
Halo mass function at redshift 0, 1 (left panel), 2 and 3 (right panel).
 We calculate the halo mass function from $N$-body simulations both with
 the force plus
 $N_p$ correction and without the correction (see text for correction in
 the details). The latter was shifted upward by a factor of 10 for clarity.
 The solid and dashed curves
 denote \citet{2011ApJ...732..122B} and \citet{1999MNRAS.308..119S}
 results, respectively.
}
\label{fig:massf}
\end{figure*}

If we want to accurately calculate the halo power spectra based on LRT,
 accurate halo mass functions are clearly needed, since the
 Lagrangian bias factor is expressed in terms of the halo mass function
 (see, Eq.~\ref{f_func}).
Here we use different mass function expressions given previously to
compare our $N$-body simulation results.
A famous numerical fit for $g(\sigma)$ is given by
\citet{1999MNRAS.308..119S} (hereafter ST) which is expressed as
\begin{equation}
 g_{\rm ST}(\sigma)=A\sqrt{\frac{2a}{\pi}}\left[1+\left(\frac{\sigma^2}{a\delta_c^2}\right)^p\right]\frac{\delta_c}{\sigma}\exp\left[-\frac{a\delta_c^2}{2\sigma^2}\right],
\end{equation}
with $A=0.3222$, $a=0.707$ and $p=0.3$.
Recently, \citet{2010MNRAS.403.1353C} (hereafter MICE) recalibrated the
halo mass function
using a large set of $N$-body simulations with good mass resolution and
large cosmological volumes called MICE simulations. 
They provided a numerical fit as
\begin{equation}
 g_{\rm MICE}(\sigma)=A(z)\left[\sigma^{-a(z)}+b(z)\right]\exp\left[-\frac{c(z)}{\sigma^2}\right],
\end{equation}
with $A(z)=0.58(1+z)^{-0.13}$, $a(z)=1.37(1+z)^{-0.15}$,
$b(z)=0.3(1+z)^{-0.084}$, and $c(z)=1.036(1+z)^{-0.024}$.
\citet{2011ApJ...732..122B} (hereafter Coyote) also recently
recalibrated the mass function using their simulations so called as Coyote
simulations whose parameters are very close to MICE simulations. 
Therefore, both results show good agreement within a few percent except
at very high masses. 
The functional form of the Coyote fit which is similar to the ST mass
function is expressed as
\begin{align}
 g_{\rm
  Coyote}(\sigma)&=\Tilde{A}\sqrt{\frac{2}{\pi}}\left[1+\left(\frac{\sigma^2}{\Tilde{a}\delta_c^2}\right)^{\Tilde{p}}\right]\nonumber\\
&\times\left(\frac{\delta_c\sqrt{\Tilde{a}}}{\sigma}\right)^{\Tilde{q}}\exp\left[-\frac{\Tilde{a}\delta_c^2}{2\sigma^2}\right],
\label{coyote_mass}
\end{align}
with $\Tilde{A}=0.333(1+z)^{-0.11}$, $\Tilde{a}=0.788(1+z)^{-0.01}$,
$\Tilde{p}=0.807$, and $\Tilde{q}=1.795$.

\begin{figure*}[!t]
\begin{minipage}{.48\textwidth}
\begin{center}
\includegraphics[width=0.95\textwidth]{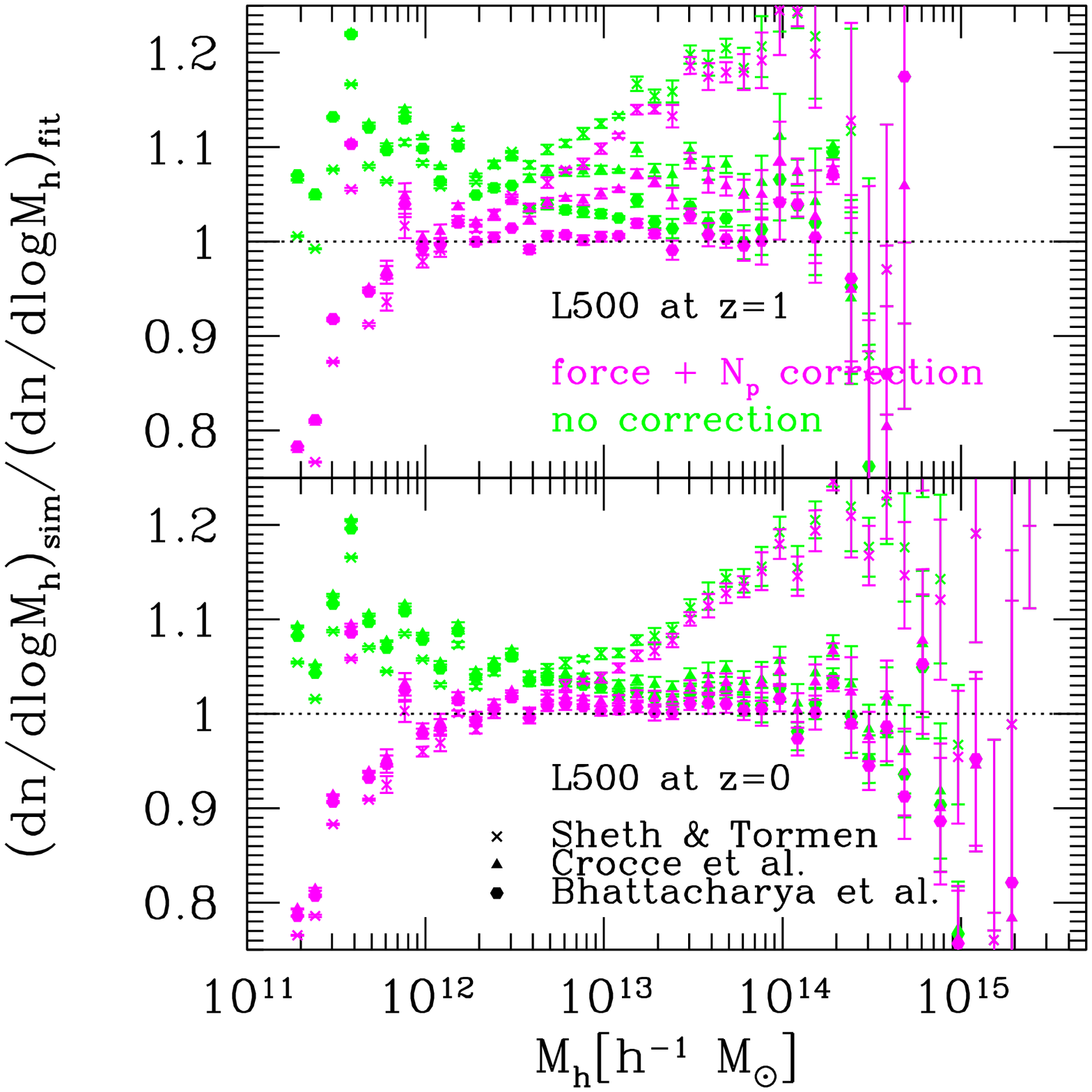}
\end{center}
\end{minipage}
\begin{minipage}{.48\textwidth}
\begin{center}
\includegraphics[width=0.95\textwidth]{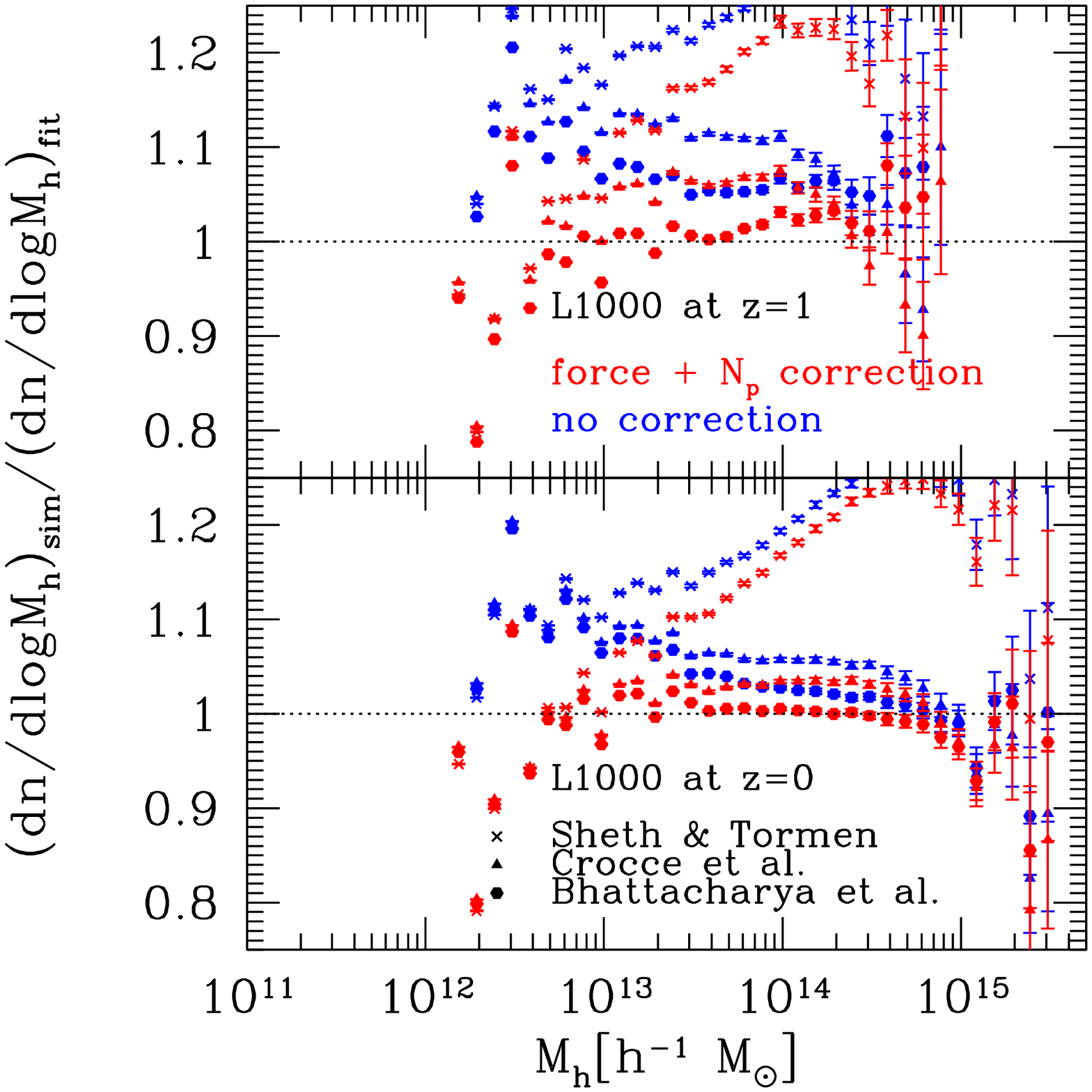}
\end{center}
\end{minipage}
\vskip-\lastskip
\caption{
 The ratio of the halo mass function from $N$-body 
 simulations to that from three fitting formulae:
 \citet{1999MNRAS.308..119S}, \citet{2010MNRAS.403.1353C},
 and \citet{2011ApJ...732..122B} for L500 (left panel) and L1000 (right panel). 
 The upper and bottom panels show the results at redshift 1 and 0, respectively.
 The red and magenta symbols show the results with the force plus $N_p$
 correction and the blue and green symbols show those without the
 correction, respectively.
}
\label{fig:massfdiff}
\end{figure*}

\begin{figure*}[!t]
\begin{minipage}{.48\textwidth}
\begin{center}
\includegraphics[width=0.95\textwidth]{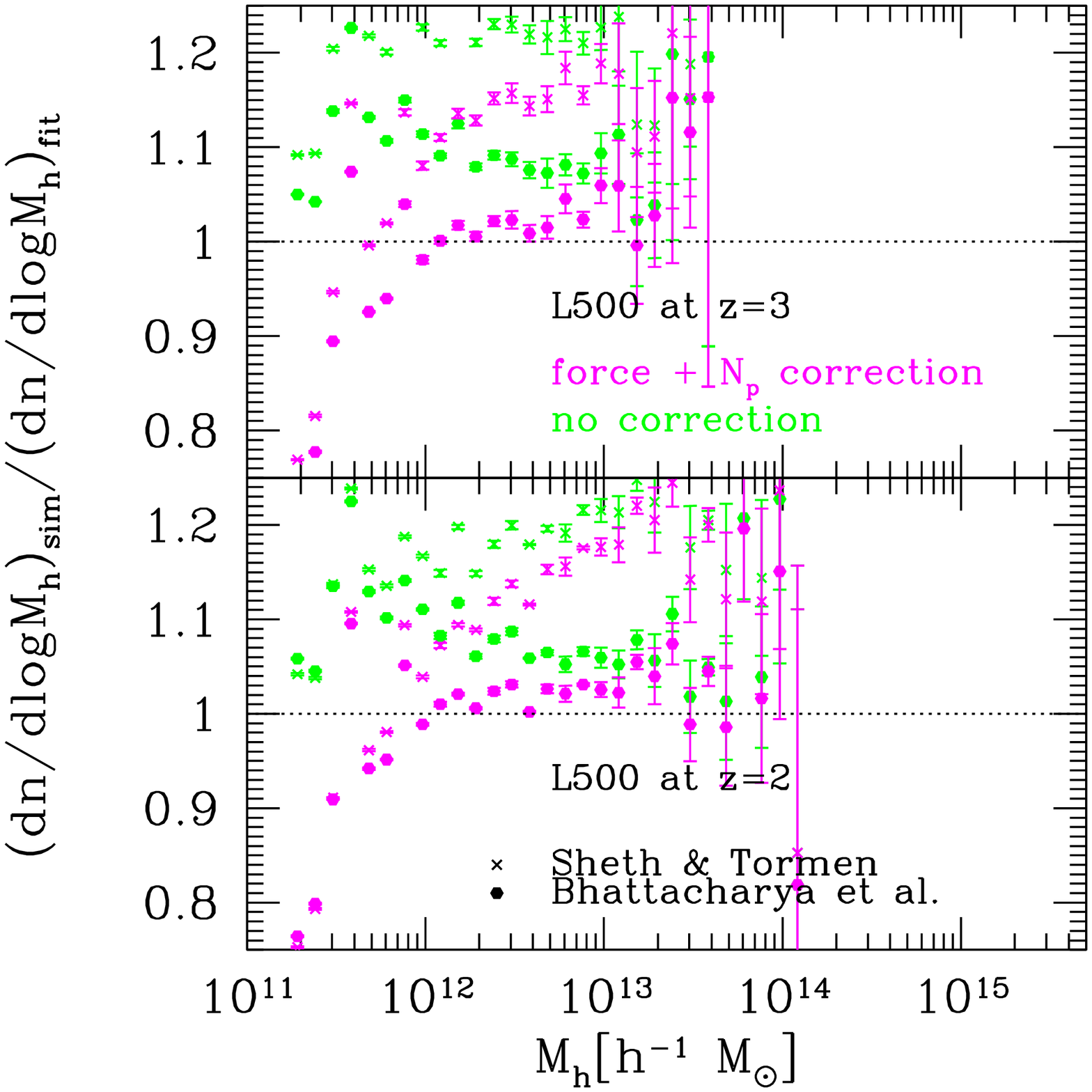}
\end{center}
\end{minipage}
\begin{minipage}{.48\textwidth}
\begin{center}
\includegraphics[width=0.95\textwidth]{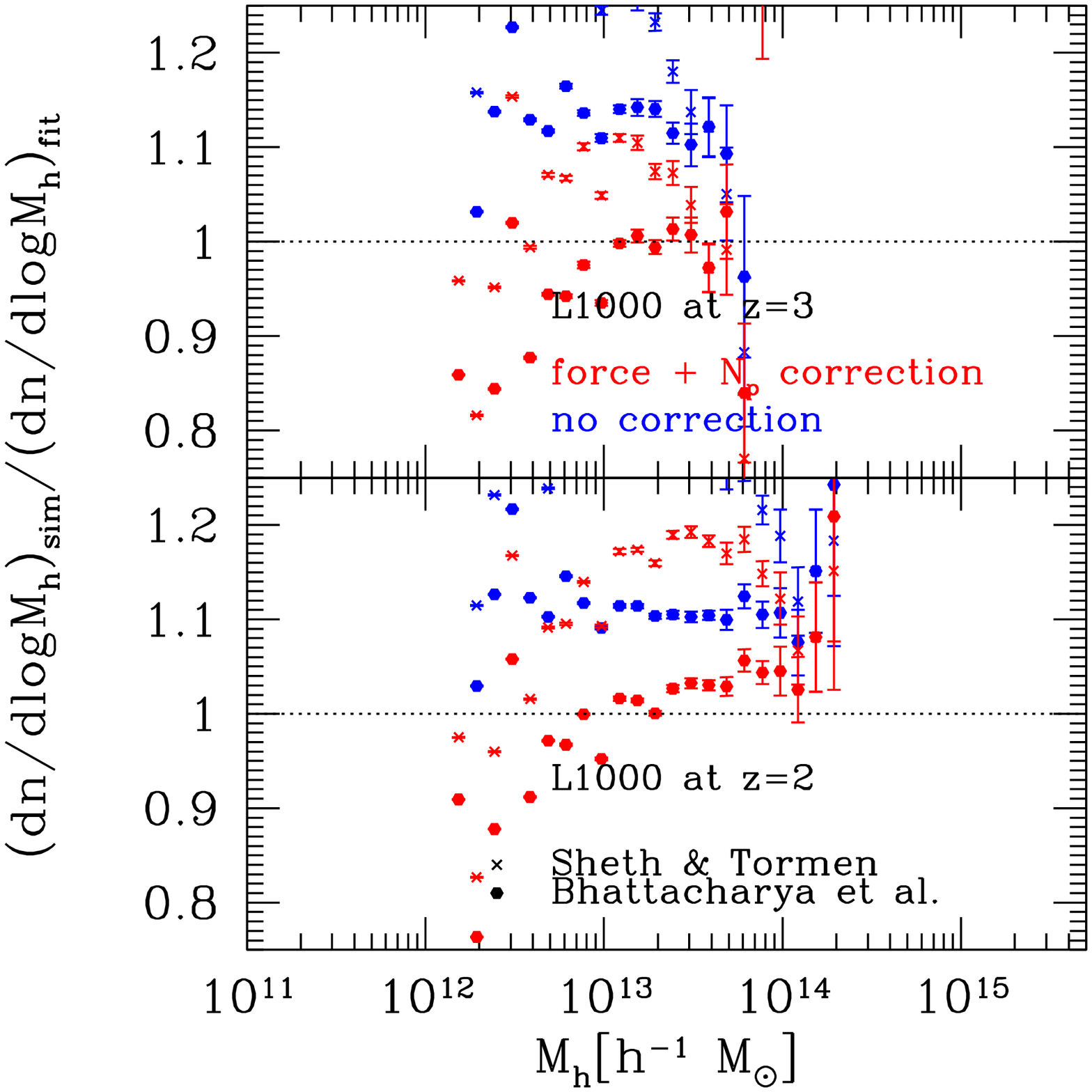}
\end{center}
\end{minipage}
\vskip-\lastskip
\caption{
 The same as Figure~\ref{fig:massfdiff}, but \citet{2010MNRAS.403.1353C}
 results are not plotted and the upper and bottom panels show the results
 at redshift 3 and 2.
}
\label{fig:massfdiff1}
\end{figure*}

In Figure~\ref{fig:massf}, we plot the halo mass function
as a function of halo mass at $z=0$, 1, 2, and 3.
We show the results of L1000 as filled symbols and the results of L500
as open symbols, respectively.
The simulation results are plotted with the force plus the $N_p$ correction
proposed by \citet{2011ApJ...732..122B} and without the correction.
The latter was shifted upwards by a factor of 10 for visibility.
The net correction for the FOF halo mass is expressed as
\citep{2011ApJ...732..122B}
\begin{equation}
 M_{h}^{c}/M_{h}=[1.0-0.04(r_s/650{\rm kpc})](1-n_{h}^{-0.65}),
\label{fofcor}
\end{equation}
where $M_{h}^{c}$ is a corrected mass, $M_{h}$ is an uncorrected mass,
and $n_{h}$ is the number of particles that construct a halos.
This equation is slightly modified from the originally suggested equation by
\citet{2006ApJ...646..881W}:
\begin{equation}
 M_{h}^{c}/M_{h}=1-n_{h}^{-0.60}.
\end{equation}
The solid and dashed curves denote the results of Coyote and ST fitting
formulae.
 Our simulation results well reproduce the Coyote
 results over a wide range of halo masses and redshifts, while
 ST slightly deviates from the simulation results, especially at large
 halo masses.
The impact of the correction to the FOF mass can be easily seen in
 Figures~\ref{fig:massfdiff} and \ref{fig:massfdiff1}.

In Figure~\ref{fig:massfdiff}, to investigate the agreement in more
quantitative ways, we show the ratio of the halo mass function obtained
from $N$-body simulations to that from fitting formulae of Coyote and
ST used in Figure~\ref{fig:massf}. In addition to the above two fitting
formulae, we use the recently proposed fitting formula by MICE. The left
and right panels show the results of L500 and L1000 at $z=0$ and 1. We
can see good agreement between our simulation results and the Coyote
fitting formula within a few percent if we apply the correction
(Eq.~\ref{fofcor}) to FOF halo masses. Also, the MICE fitting formula
reproduces the simulation results applying FOF mass correction with
the same level in Coyote at $z=0$, although the result at $z=1$
deviates about 5$\%$ from our simulation result. Meanwhile, ST
results underpredict the simulation results by about 10$\%$-20$\%$ at
$M_h>10^{13}h^{-1}M_{\odot}$.
Therefore, the halo mass function obtained from our $N$-body
simulations supports recently proposed fitting formulae obtained by
using large and high-resolution $N$-body simulations and vice versa.

However, a closer look at $M_h<10^{12}h^{-1}M_{\odot}$ in the left panel and
$M_h<10^{13}h^{-1}M_{\odot}$ in the right panel reveals deviation from unity
which shows that the FOF correction may not be perfect in halos with small
particles.
Therefore, an accurate mass estimation requires keeping many more particles
within individual halos, although halos with only a small number of
particles ($\sim 20$) can be found using the FOF algorithm.
Or a more accurate FOF mass correction needs to be developed. 
In addition, aside from simple considerations of particle shot noise,
there is an inherent systematic error and scatter in
the definition of a FOF halo mass with particle number, as pointed out
by \citet{2006ApJ...646..881W}.

Figure~\ref{fig:massfdiff1} is the same as
Figure~\ref{fig:massfdiff}, but at redshift 2 and 3 and results of
the MICE fitting formula are not plotted because their fitting formula was
fitted in the redshift range $z=0-1$. 
As in the results at $z=0$ and 1, our results are in good agreement
with the Coyote results, although the application of
their fitting formula to the result at $z=3$ might not be reliable, because
the Coyote fitting formula was fitted between $z=0$ and $z=2$.
As shown in Figure~\ref{fig:massfdiff}, the deviation from unity at low
mass halos is also shown in Figure~\ref{fig:massfdiff1}.



\subsection{Halo Biases}

\begin{figure*}[!t]
\begin{minipage}{.48\textwidth}
\begin{center}
\includegraphics[width=0.95\textwidth]{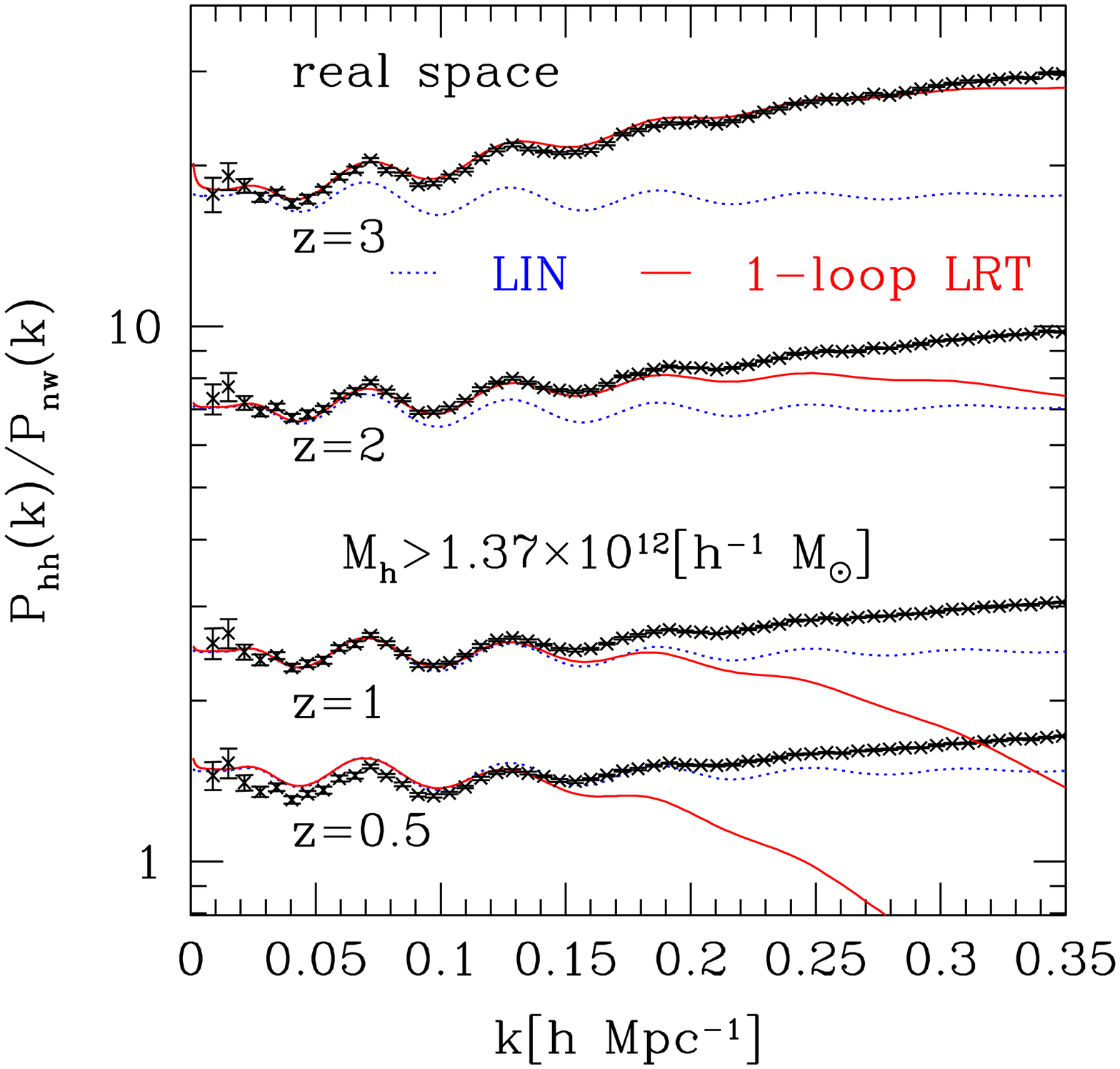}
\end{center}
\end{minipage}
\begin{minipage}{.48\textwidth}
\begin{center}
\includegraphics[width=0.95\textwidth]{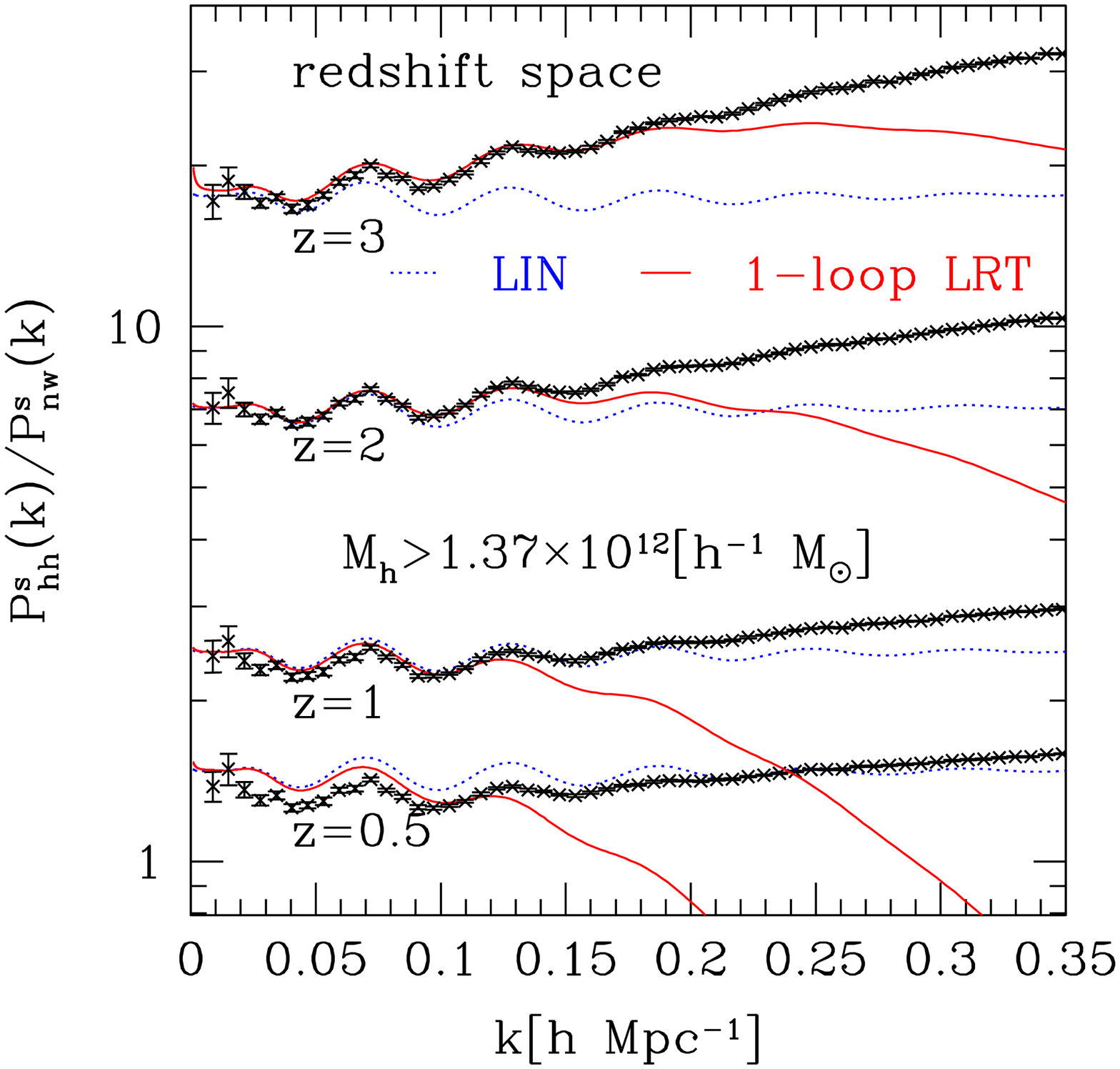}
\end{center}
\end{minipage}
\vskip-\lastskip
\caption{
 Comparison of halo power spectra obtained from $N$-body
 simulations to analytical predictions in real (left panel) and redshift
 (right panel) space for redshifts, $z=3$, 2, 1, and 0.5.
 Halo power spectra are normalized by the no-wiggle fitting formula $P_{\rm
 nw}(k)$ \citep{1998ApJ...496..605E} in real space and by the no-wiggle
 fitting formula taking the Kaiser effect into account $P^s_{\rm nw}(k)$ in
 redshift space, respectively. Therefore, the vertical axis shows the
 square of the halo bias.
 The solid and dotted curves show the predictions of the 1-loop Lagrangian
 resummation theory and linear theory, respectively.
 An enhancement of amplitude on small scales is due to the nonlinear effects of
 dynamics, bias, and redshift-space distortions.
}
\label{fig:halobias}
\end{figure*}

In Figure~\ref{fig:halobias} we compare the halo power spectra
computed from $N$-body simulations with the analytical predictions in real
(left panel) and redshift (right panel) space for $z=3$, 2, 1, and
0.5. Halo power spectra are normalized by the no-wiggle fitting formula $P_{\rm
  nw}(k)$ \citep{1998ApJ...496..605E} in real space and by the
no-wiggle fitting formula taking the Kaiser effect into account $P^s_{\rm
  nw}(k)$ in redshift space, respectively. In the monopole
power spectrum we have considered, Kaiser's enhancement effect $R_{\rm
  hh}$ is expressed as
\begin{equation}
 R_{\rm hh}=1+\frac{2}{3}\beta+\frac{1}{5}\beta^2,
\label{kaiser_hh}
\end{equation}
from an angular average of the factor $(1+\beta\mu^2)^2$ in
Eq.~(\ref{P_HLIN_R}). Therefore
the vertical axis corresponds to the square of the halo bias, i.e.,
$b^2(k)$. The solid and dotted curves show the predictions of 1-loop LRT
(Eqs.~\ref{P_HLRT_R} and \ref{P_HLRT}) and LIN (Eqs.~\ref{P_HLIN_O}
and \ref{P_HLIN_R}), respectively. When we calculate the prediction of the
1-loop LRT, we use the Coyote fit (Eq.~\ref{coyote_mass}) for the halo
mass function
because the prediction of the Coyote fit is well duplicated by our simulation
results as seen in Figures~\ref{fig:massfdiff} and
\ref{fig:massfdiff1}.

As increasing redshift, the amplitude of power spectrum becomes larger
because we impose the same halo mass threshold ($M_{\rm h}>1.37\times
10^{12}h^{-1}M_{\odot}$) regardless of redshift. Halos which have a
certain mass, for example, this threshold mass $M_{\rm h}>1.37\times
10^{12}h^{-1}M_{\odot}$, are increasingly rare as increasing redshift,
and thus the amplitude of the power spectrum is more biased. Theoretical
predictions of 1-loop LRT and LIN well agree with $N$-body simulations at
large scales up to a certain scales. As in the matter power spectrum, the
range of agreement in the 1-loop LRT prediction is wider as redshift is higher.
Meanwhile the LIN prediction significantly deviates from the $N$-body results
as bias is higher. Therefore, the 1-loop LRT prediction which include
higher-order bias term is very powerful for future high redshift BAO surveys to
extract the cosmological information. At very large scales if the mass
function prediction is accurate, 1-loop LRT and LIN predictions should
reproduce the simulation result because all the higher-order
Lagrangian bias factors $\langle{F^{(n)}}\rangle$ should be zero except
for $\langle{F'}\rangle$. However, a closer look at small wave number $k$
for $z=0.5$ reveals a discrepancy in overall amplitudes between
theoretical predictions and simulation results beyond the range
of error bars. This might be ascribed to the fact that the FOF
mass correction (Eq.~\ref{fofcor}) is not perfect in halos with small
particles as shown in Figures~\ref{fig:massfdiff} and
\ref{fig:massfdiff1}.

\begin{figure*}[!t]
\begin{minipage}{.48\textwidth}
\begin{center}
\includegraphics[width=0.95\textwidth]{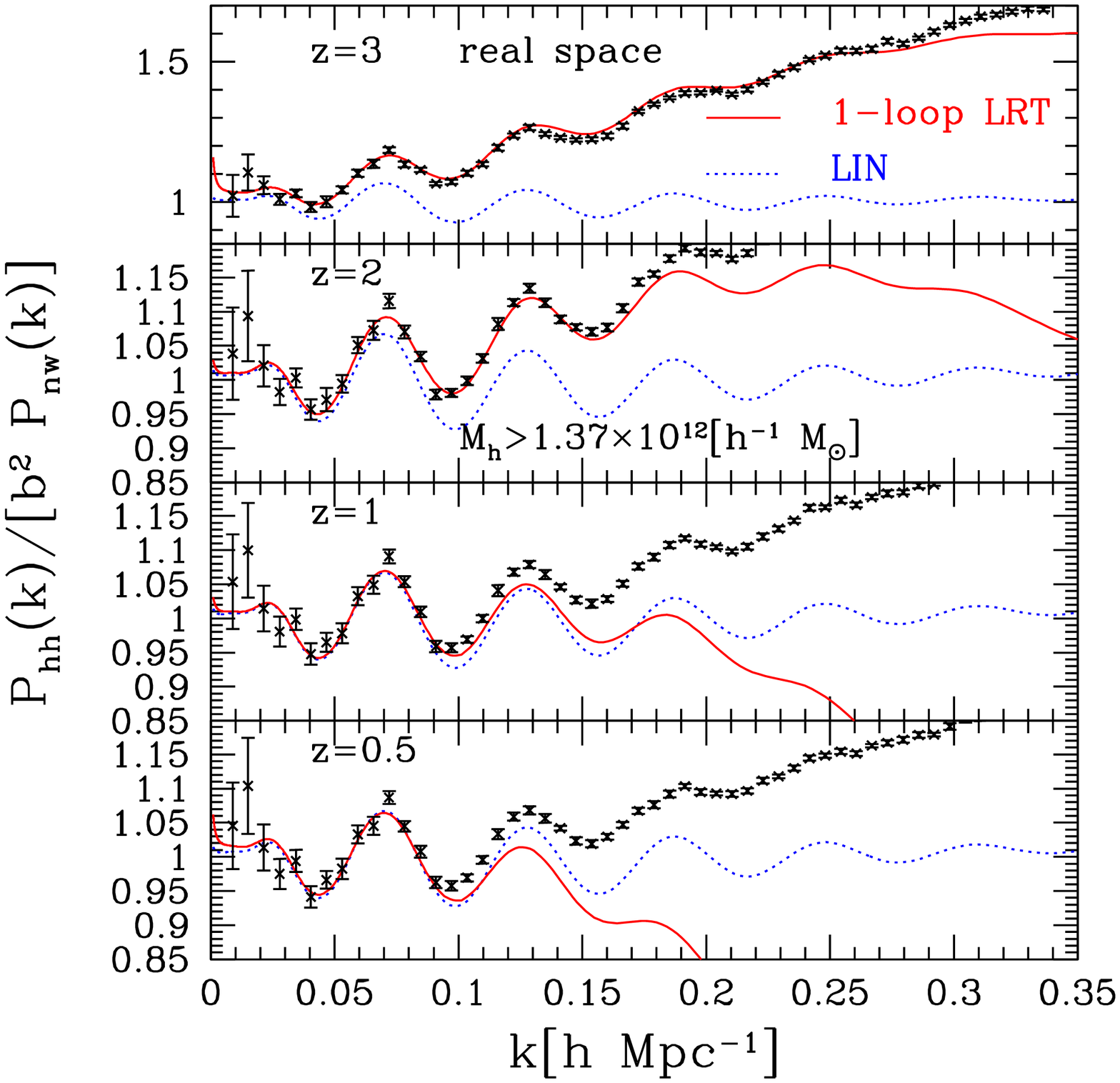}
\end{center}
\end{minipage}
\begin{minipage}{.48\textwidth}
\begin{center}
\includegraphics[width=0.95\textwidth]{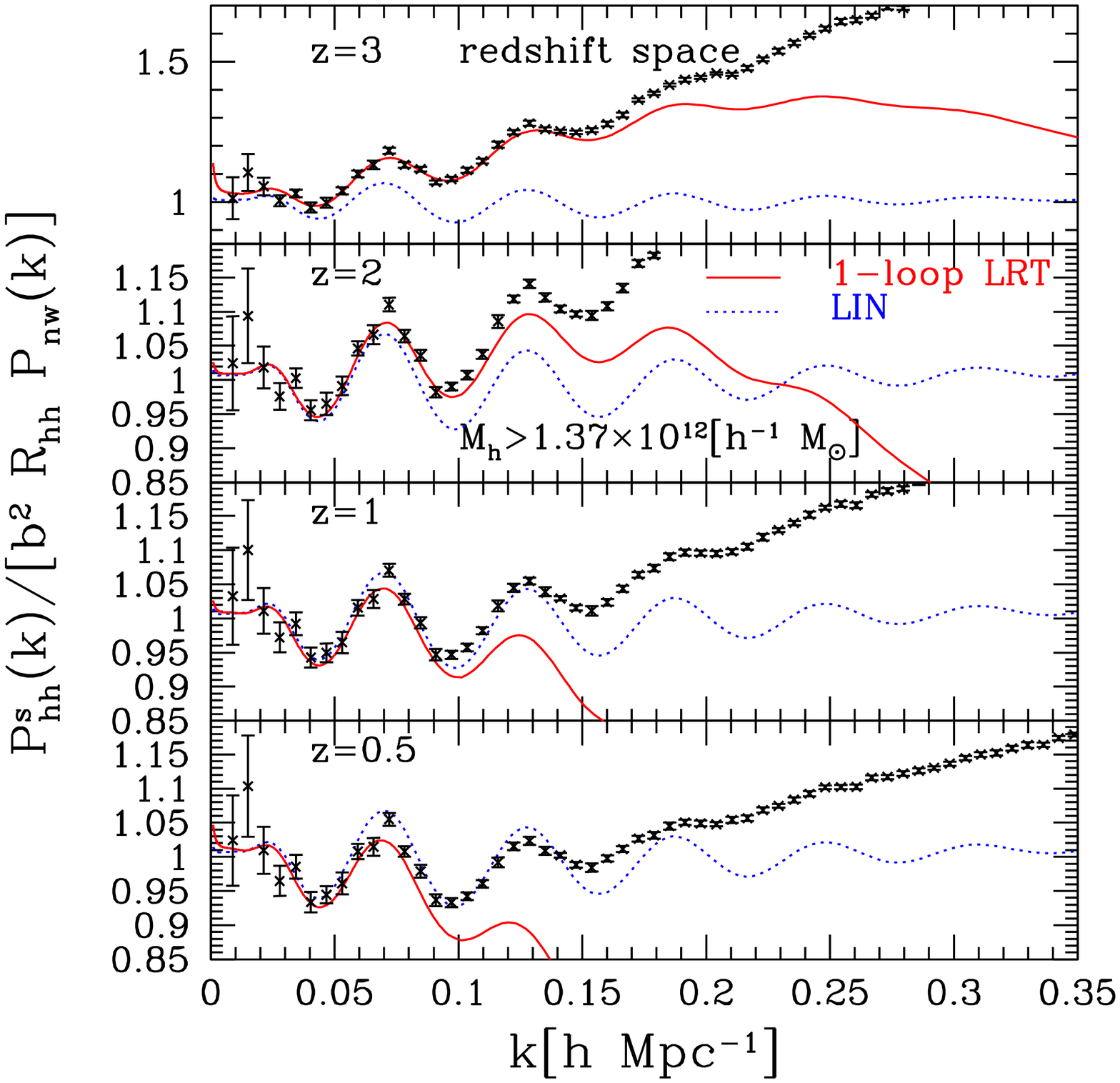}
\end{center}
\end{minipage}
\vskip-\lastskip
\caption{
The same as Figure~\ref{fig:halobias}, but the vertical axis is divided by
 $b^2=(1+\langle{F'}\rangle)^2$
 and $b^2R_{\rm hh}=(1+\langle{F'}\rangle)^2(1+2\beta/3+\beta^2/5)$
 in real and redshift space, respectively.
 An enhancement of amplitude on small scales comes from the nonlinear effects of
 dynamics, bias, and redshift-space distortions.
}
\label{fig:halopk}
\end{figure*}

In Figure~\ref{fig:halopk}, to examine the agreement in more
quantitative ways and exclude an uncertainty in amplitudes
explained above, we eliminate dependences on the linear bias
factor and linear Kaiser's factor, dividing the power spectra by
$b^2=(1+\langle{F'}\rangle)^2$ and $b^2R_{\rm
  hh}=(1+\langle{F'}\rangle)^2(1+2\beta/3+\beta^2/5)$ in real and
redshift space, respectively. For simulation results $b^2$ and
$b^2R_{\rm hh}$ are numerically estimated as follows.
First, those factors can be computed among 30 realizations
through
\begin{equation}
 b^2(k)=\frac{P_{\rm hh}(k)}{P_{\rm mm}(k)},\qquad b^2(k)R_{\rm hh}=\frac{P_{\rm
  hh}^{\rm s}(k)}{P_{\rm mm}(k)},
\end{equation}
where $P_{\rm mm}(k)$ denotes the real-space power spectrum of matter,
 and $P_{\rm hh}(k)$ and $P_{\rm hh}^{\rm s}(k)$ denote the halo power
 spectrum in real and redshift space. Those power spectra are obtained from
$N$-body simulations. Second, we use polynomial fitting with
\begin{align}
 b^2(k)&=A_{0}+\sum_{i=1}^{3}A_{i}k^{i},\\
 b^2(k)R_{\rm hh}&=B_{0}+\sum_{i=1}^{3}B_{i}k^{i}.
\end{align}
Finally, computing the $\chi^2$ statistics with the data up to
$k$=0.35$h$Mpc$^{-1}$,
we can obtain the best-fit values of scale independent terms,
$A_0$ and $B_0$, and then we substitute these values 
into $b^2$ and $b^2R_{\rm hh}$, respectively.

\begin{figure*}[!t]
\begin{minipage}{.48\textwidth}
\begin{center}
\includegraphics[width=0.95\textwidth]{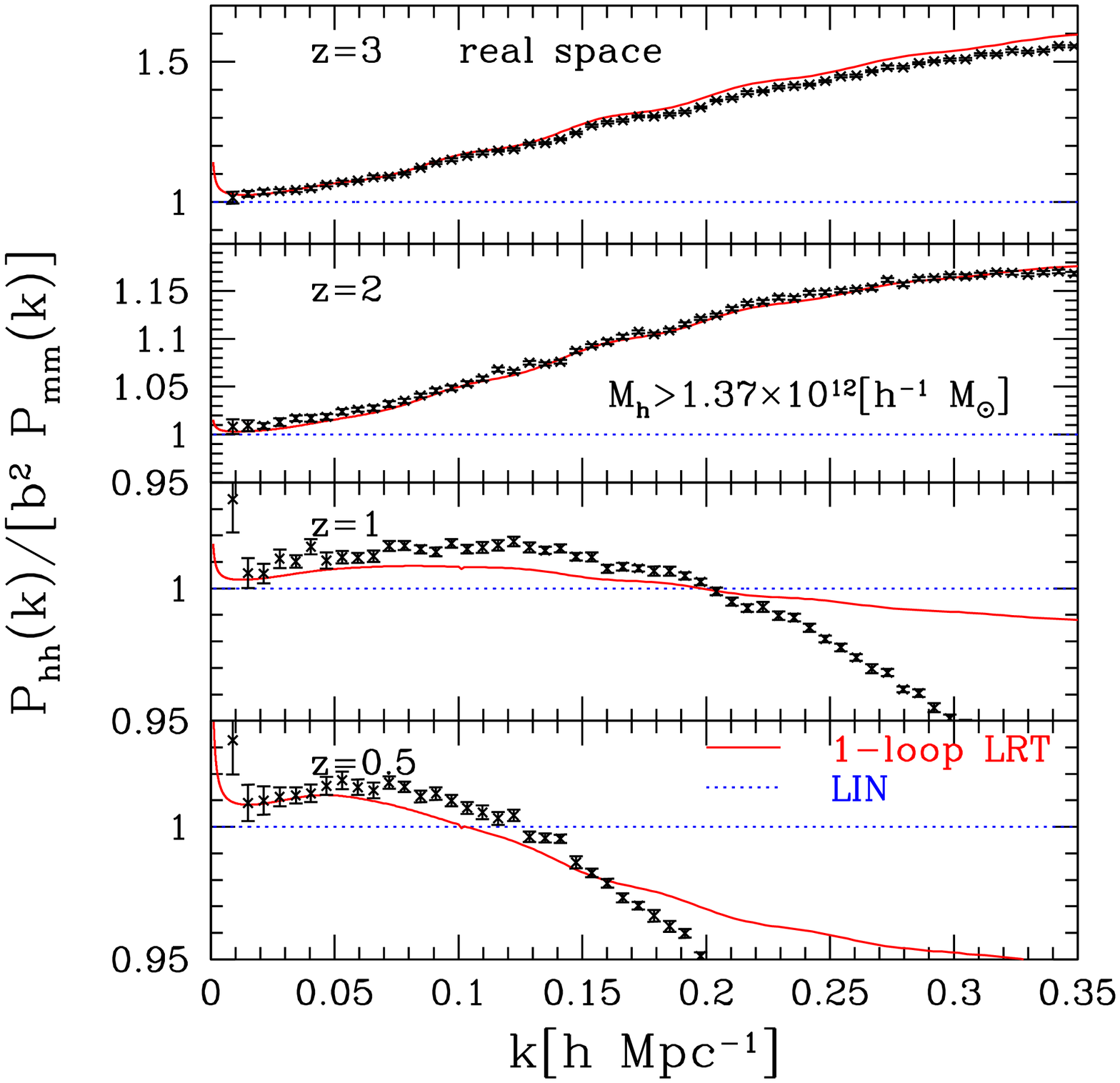}
\end{center}
\end{minipage}
\begin{minipage}{.48\textwidth}
\begin{center}
\includegraphics[width=0.95\textwidth]{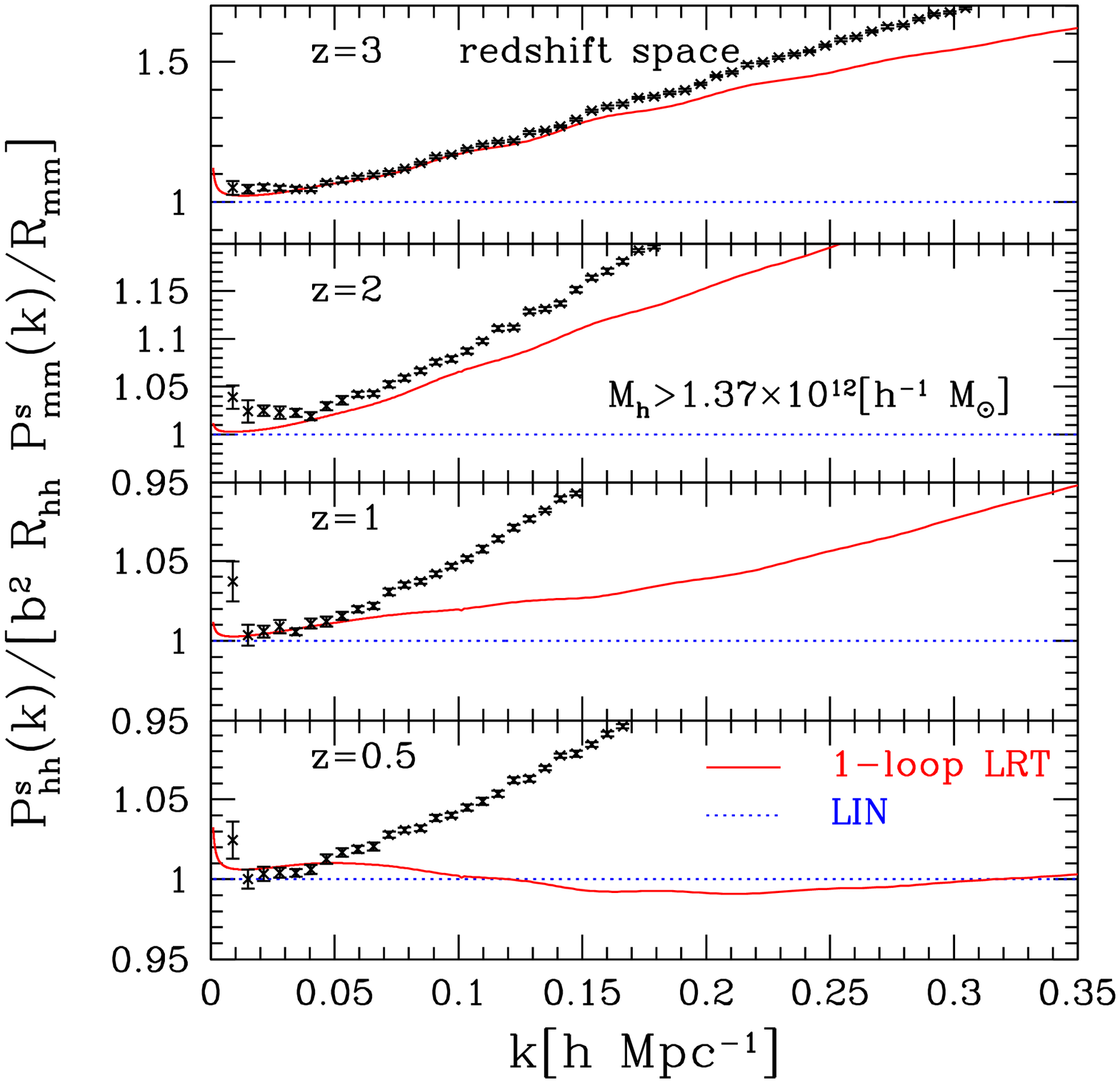}
\end{center}
\end{minipage}
\vskip-\lastskip
\caption{
The same as Figure~\ref{fig:halopk}, but the smooth linear power spectrum
 $P_{\rm nw}(k)$ is replaced with $P_{\rm mm}(k)$ and $P_{\rm mm}^{\rm
 s}(k)/R_{\rm mm}$ in real and redshift space, respectively in order to
 get rid of the nonlinearity of dynamics.
 Therefore deviation from unity shows the nonlinearity of bias and that
 of bias and redshift-space distortions in real and redshift space.
 $P_{\rm mm}(k)$ and $P_{\rm mm}^{\rm s}(k)$ are corresponding mass power
 spectra in real and redshift space.
 $R_{\rm mm}$ is Kaiser's enhancement factor of matter defined by $R_{\rm
 mm}=1+2f/3+f^2/5$.
}
\label{fig:scalebias}
\end{figure*}

Comparing the matter power spectra in Figure~\ref{fig:matter_pk} and
halo power spectra in Figure~\ref{fig:halopk}, the ranges of
agreement between $N$-body simulations and 1-loop LRT predictions with halo
bias seem to be the same as those in the matter power spectrum for all redshift
we have examined in real space. However, the results of the halo power
spectrum in redshift space seem to be somewhat worse than those of
matter power spectrum as decreasing redshift. This might be because
the effects of nonlinear redshift-space distortions become larger as
decreasing redshift. The redshift-space distortion parameter $\beta
\simeq \Omega_m^{0.55}/b$ becomes larger
as decreasing redshift since a halos at lower redshift are less
biased than those at higher redshift. Therefore, 1-loop LRT well replicate
the $N$-body simulation results on BAO scales even in existence of
halo bias, although the results of low redshifts in redshift
  space are less accurately reproduced by 1-loop LRT. Comparing the results
of real space at $z=0.5$ and $z=1$ in Figure~\ref{fig:matter_pk} and
those in Figure~\ref{fig:halopk}, an enhancement of amplitude of halo
power spectra on small scales is slightly moderate because
$\langle{F''}\rangle$ has a negative value, which is calculated from
Eq.~(\ref{f_func}), where we substitute the minimum halo mass in our
simulations, $M_{\rm h}=1.37\times 10^{12}h^{-1}M_{\odot}$, into $M_1$
and infinity into $M_2$. In contrast, other redshift results ($z=2$
and 3) of halos in real space are more enhanced compared to those of
matter power spectrum, since $\langle{F''}\rangle$ has a positive
value at $z=2$ and 3. In redshift space, amplitudes of the results of
halo power spectrum are enhanced compared to those of matter power
spectrum at all redshift. This is due to the nonlinear effects of bias
and redshift-space distortions. This scale dependence of bias coming
from the clustering of halos and nonlinear redshift-space distortions
will be carefully examined in Figure~\ref{fig:scalebias}.


Figure~\ref{fig:scalebias} is the same as Figure~\ref{fig:halopk}, but
the smooth no-wiggle linear power spectrum $P_{\rm nw}(k)$ is replaced with
$P_{\rm mm}(k)$ and $P_{\rm mm}^{\rm s}(k)/R_{\rm mm}$ in real and
redshift space, respectively.
Here $P_{\rm mm}(k)$ and $P_{\rm mm}^{\rm s}(k)$ are corresponding mass
power spectra in real and redshift space.
$R_{\rm mm}$ is Kaiser's enhancement factor of matter, which is obtained
from Eq.~(\ref{kaiser_hh}) with $b=1$ as 
\begin{equation}
 R_{\rm mm}=1+\frac{2}{3}f+\frac{1}{5}f^2.
\end{equation}
For simulation results in redshift space, $b^2R_{\rm hh}/R_{\rm
mm}$ is estimated from simulations through
\begin{equation}
b^2(k)R_{\rm hh}/R_{\rm mm}=\frac{P_{\rm hh}^{\rm s}(k)}{P_{\rm mm}^{\rm
 s}(k)}.
\end{equation}
Using polynomial fitting with
\begin{equation}
b^2(k)R_{\rm hh}/R_{\rm mm}=C_{0}+\sum_{i=1}^{3}C_{i}k^{i},
\end{equation}
we calculate the $\chi^2$ statistics with the data up to
$k$=0.35$h$Mpc$^{-1}$ and search for 
the best-fit value of scale independent term, $C_0$. 
Then, we substitute the best-fit value into $b^2R_{\rm hh}/R_{\rm mm}$.

The functional form of vertical axis in real space, $P_{\rm
  hh}(k)/[b^2 P_{\rm mm}(k)]$, describes only the nonlinear effect of
bias, so a deviation from unity shows the scale dependence of
bias. Meanwhile in redshift space the vertical axis shows nonlinear
effects of bias and redshift-space distortions. We find that the scale
dependence of bias does not show significant oscillations and are
mostly smooth functions of scales. The scale dependence and
nonlinearity of bias shown in our results purely originate from
clustering of halos themselves. LIN predicts constancy of halo bias
on all scales. We find that scale dependences of bias calculated from
our simulation results are in fairly good agreement with 1-loop LRT
predictions up to $k=0.35h$Mpc$^{-1}$ ($z=2$ and 3) within a few percent
level and up to $k=0.1h$ Mpc$^{-1}$ ($z=2$) and 0.15$h$Mpc$^{-1}$ ($z=3$) in
redshift space. Recent analytical and numerical studies claim that the
scale-dependent and stochastic properties of the bias can change the
redshift-space power spectrum and those impacts on the determination
of the growth-rate parameter would be significant
\citep{2009ApJ...691..569J,2010PhRvD..81b3526D,2011MNRAS.410.2081J,2011ApJ...726....5O,2011PhRvD..83d3529S}.
Therefore LRT prediction is useful and powerful to model biasing
relevant for BAO scales.

\begin{figure*}[!t]
\begin{minipage}{.48\textwidth}
\begin{center}
\includegraphics[width=0.95\textwidth]{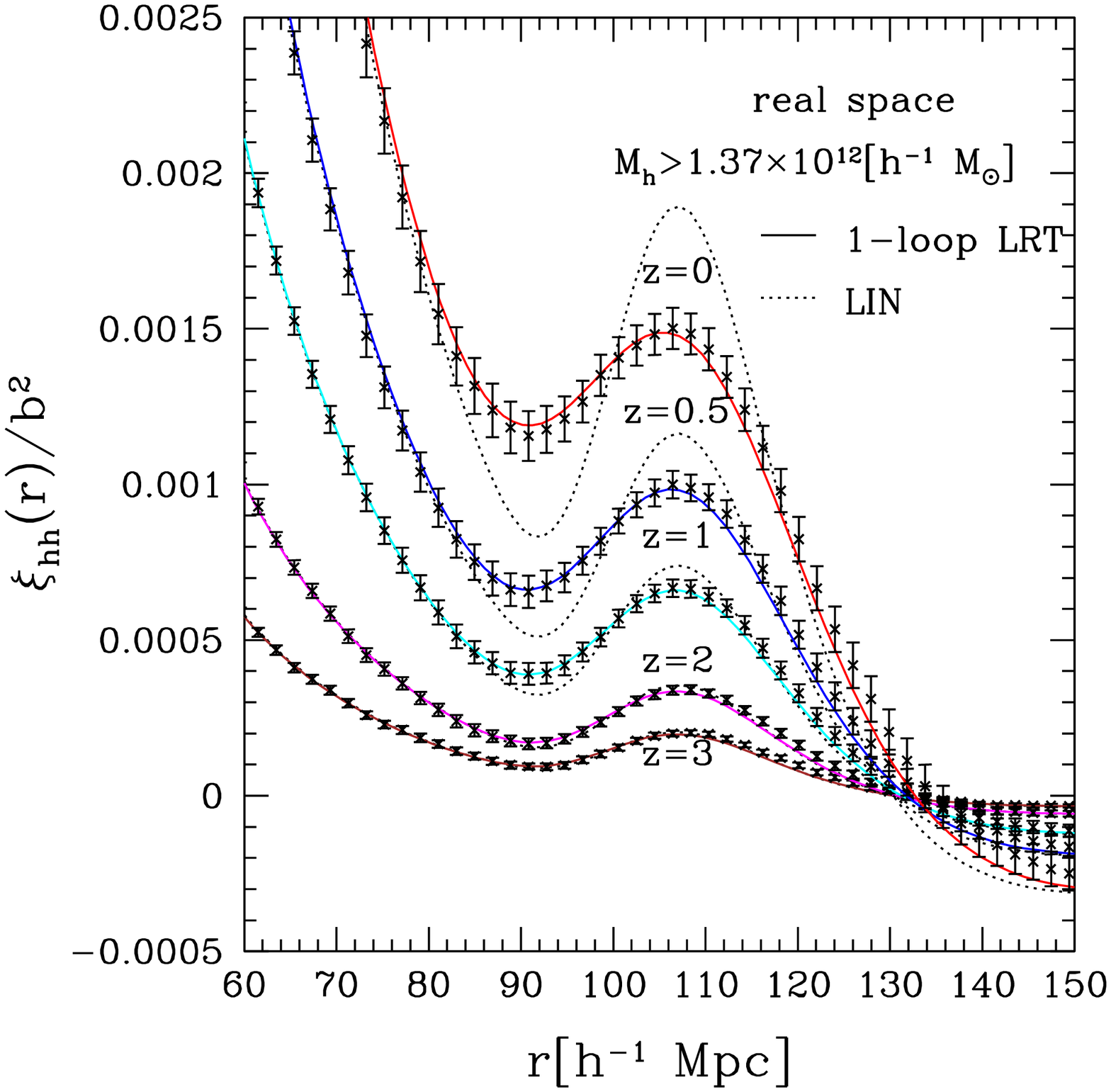}
\end{center}
\end{minipage}
\begin{minipage}{.48\textwidth}
\begin{center}
\includegraphics[width=0.95\textwidth]{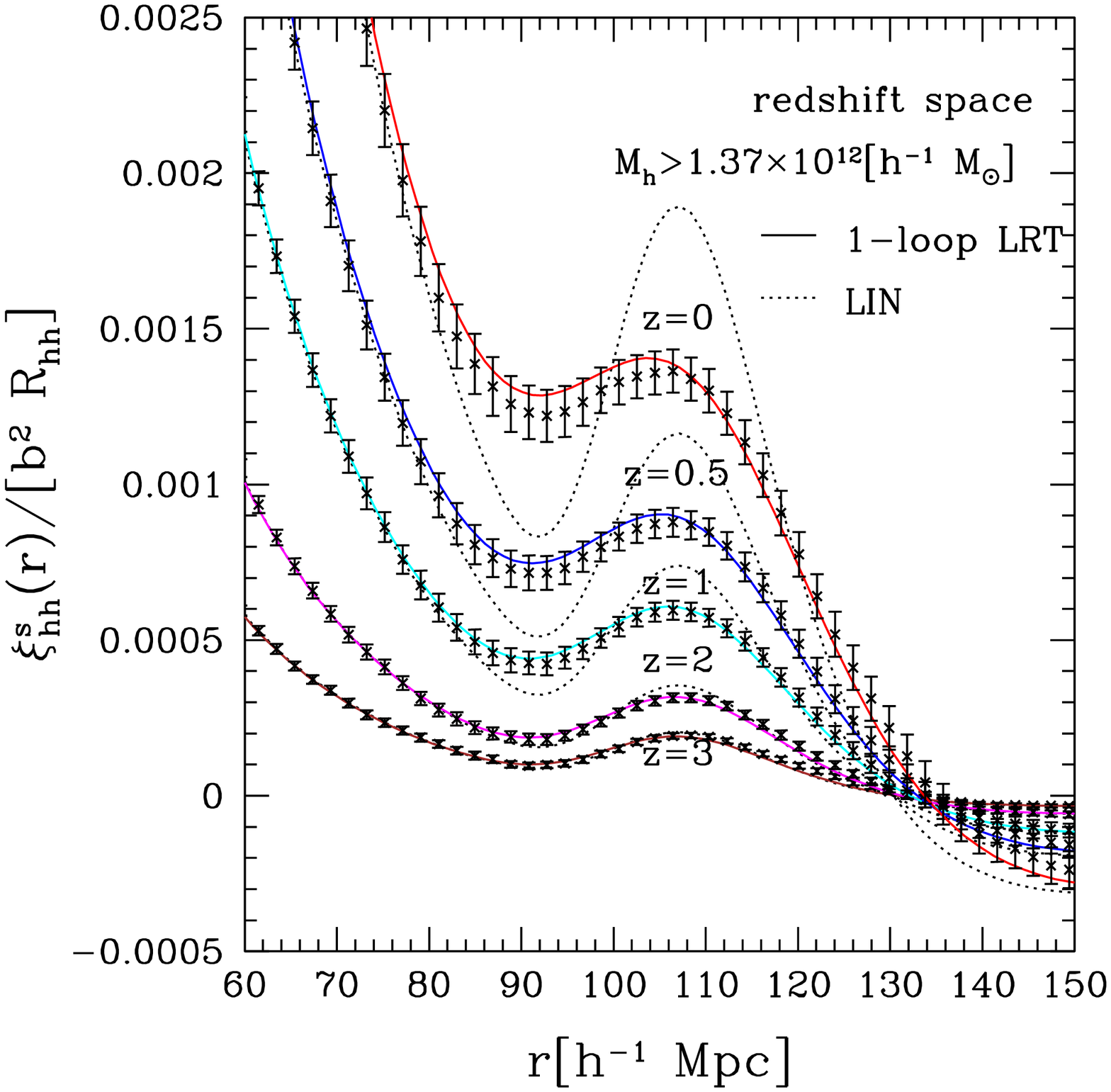}
\end{center}
\end{minipage}
\vskip-\lastskip
\caption{
Comparison of halo two-point correlation functions computed from
 $N$-body simulations to analytical predictions in real (left panel) and
 redshift (right panel) space for redshifts, $z=3$, 2, 1, 0.5 and 0 from
 bottom to top.
 Halo correlation functions are normalized by $b^2=(1+\langle{F'}\rangle)^2$
 and $b^2R_{\rm hh}=(1+\langle{F'}\rangle)^2(1+2\beta/3+\beta^2/5)$
 in real and redshift space, respectively.
 The solid curves represent the results of the 1-loop Lagrangian
 resummation theory while the dotted curves show the results of the linear
 theory.
}
\label{fig:haloxi}
\end{figure*}

Finally, we examine the two-point correlation function of halos for
$z=3$, 2, 1, 0.5, and 0 both in real and redshift space, shown in the left
and right panels of Figure~\ref{fig:haloxi}, respectively.
Two-point correlation functions of halos are divided by
$b^2=(1+\langle{F'}\rangle)^2$ and $b^2R_{\rm
  hh}=(1+\langle{F'}\rangle)^2(1+2\beta/3+\beta^2/5)$ in real and
redshift space, respectively. The solid curves represent the results
of 1-loop LRT while the dotted curves show the results of LIN. For
simulation results $b^2$ and $b^2R_{\rm hh}$ are fitted in the
  same manner as in the case of the halo power spectrum
(Figures~\ref{fig:halopk} and \ref{fig:scalebias}). 
Those can be calculated among 30 realizations through
\begin{equation}
 b^2(r)=\frac{\xi_{\rm hh}(r)}{\xi_{\rm mm}(r)},\qquad b^2(r)R_{\rm
  hh}=\frac{\xi_{\rm hh}^{\rm s}(r)}{\xi_{\rm mm}(r)},
\end{equation}
where $\xi_{\rm mm}(r)$ denotes the matter correlation function in real
space, and $\xi_{\rm hh}(r)$ and $\xi_{\rm hh}^{\rm s}(r)$ denote the halo
correlation functions in real and redshift space. Those correlation
functions are obtained from $N$-body simulations.
We assume $b^2(r)$ and $b^2(r)R_{\rm hh}$ to be constant
on sufficiently large scales both in real and redshift space and chose
$30h^{-1}$Mpc $<r<80h^{-1}$Mpc in configuration space. Then we compute
$\chi^2$ for the above range to obtain $b^2$ and $b^2R_{\rm hh}$.

The left panel in Figure~\ref{fig:haloxi} is quite similar to the left
panel in Figure~\ref{fig:matter_xi}. This shows that halo bias does
not significantly change the shape of BAOs peak and that the effects
of nonlinear halo bias are not significantly relevant for BAO scales
in the correlation function. This result is consistent with a recent
analysis of halo clustering by using numerical simulations
\citep{2008MNRAS.390.1470S}. By comparing the left and right
panels in Figure~\ref{fig:haloxi}, we can see the nonlinear effect of
redshift-space distortions. It smears the BAO peaks and troughs as 
decreasing the redshift. One-loop LRT with halo bias shows good
agreement with $N$-body simulation results and well captures the
features of nonlinear effects as in the matter correlation
function. 
The results of 1-loop LRT in redshift space slightly deviate from those
of simulations at low redshifts.
This should be related to the inaccuracy of the 1-loop LRT in the low-redshift
power spectra as seen in Figure~\ref{fig:halopk}.


\section{Conclusion}
\label{sec:conc}
To exploit the full potential of upcoming high-quality data, we have
to precisely describe observable features of BAOs. To
  achieve this purpose, taking account of the galaxy biasing and
  redshift-space distortions is essential. In this paper, we have
used 30 large cosmological $N$-body simulations of the standard
$\Lambda$CDM cosmology to investigate the halo biases over a wide
redshift range.

First, in the matter power spectrum, 1-loop LRT is useful for studying
nonlinear effects on scales of BAOs and provides better
  agreement with $N$-body results than 1-loop SPT or the prediction of
\citet{2004PhRvD..70h3007S} on large scales in observable redshift
space. Second, in the matter correlation function, 1-loop LRT prediction
well describes the acoustic peaks and nonlinear smearing effects
both in real and redshift space. The 2-loop correction to LRT
generally extends the valid range in the matter power spectrum. It does
not have much impact on the correlation function in real space, because
1-loop LRT is already accurate enough to describe the nonlinear effects on BAO
scales in $N$-body simulations \citep{2011arXiv1105.1491O}. 
  However, the predictions of 1-loop LRT for low-redshift correlation
  functions in redshift space would be improved if the 2-loop
  corrections are included.

  In the halo power spectrum, we found that 1-loop LRT prediction has good
  agreement with $N$-body simulation results. The ranges of agreement
  seem the same as those in the matter power spectrum for all redshift in
  real space. In redshift space, the 1-loop LRT prediction for the halo power
  spectrum is slightly worse than that for the matter power spectrum. 
  This might be because the nonlinear effects of redshift-space
  distortions become larger as decreasing the redshift.
  This shows that the 1-loop LRT prediction well reproduces the $N$-body simulation
  results on BAO scales even in the existence of halo bias at least in
  real space. We found that the scale dependences of bias are pretty
  well reproduced by 1-loop LRT up to $k=0.35h$Mpc$^{-1}$ ($z=2$ and 3)
  within a few percent level in real space and up to $k=0.1h$Mpc$^{-1}$
  ($z=2$) and 0.15$h$Mpc$^{-1}$ ($z=$3) in redshift space. In the halo correlation
  function, 1-loop LRT well describes nonlinear effects that smear the
  baryon peak and trough both in real and redshift space. Halo bias
  does not significantly change the shape of the baryon peak. Therefore,
  the nonlinear effects of halo bias are not serious on BAO scales.

Thus, LRT prediction is very powerful and reliable to accurately extract
cosmological information for upcoming high redshift BAO surveys.

\acknowledgments

We greatly appreciate Takahiro Nishimichi for kindly providing
parallelized 2nd-order Lagrangian perturbation theory code.
M.S. is supported by a Grant-in-Aid for the Japan Society for Promotion of
Science (JSPS) fellows. T.M. acknowledges support from the
Ministry of Education, Culture, Sports, Science, and Technology (MEXT),
Grant-in-Aid for Scientific Research (C), No.~21540263, 2009. This
work is supported in part by the JSPS Core-to-Core Program ``International
Research Network for Dark Energy'' and by a Grant-in-Aid for
Scientific Research on Priority Areas No. 467 ``Probing the Dark
Energy through an Extremely Wide and Deep Survey with Subaru
Telescope'' and by a Grant-in-Aid for Nagoya University Global COE
Program, ``Quest for Fundamental Principles in the Universe: from
Particles to the Solar System and the Cosmos'', from the MEXT of
Japan.


\bibliography{ms}

\clearpage

\end{document}